\documentclass[11pt,reqno]{amsart}
\usepackage{amssymb,latexsym}
\usepackage{url}
\usepackage[dvipdfm]{color}
\usepackage{graphicx, amsmath, amssymb}
\usepackage{epsfig}
\usepackage[pdfstartview=FitH,pdfpagemode=None,backref,colorlinks=true,citecolor=bluetxt,linkcolor=bluetxt]{hyperref} 

\newcommand{\set}[1]{\left\{#1\right\}}

\newcommand{\eps}{\varepsilon}

\newcommand {\cd}{\cdot}

\newcommand {\ind} {\noindent}

\newcommand{\SmallO}{o}

\def\ps0{polynomial-size}
\newcommand{\mbf}{\mathbf}

\newcommand{\NP}{\mbf{NP}}
\newcommand{\coNP}{\textbf{co-NP}}

\renewcommand{\Im}{\mathrm{Im}}

\theoremstyle{plain}
\newtheorem{theorem}{Theorem}
\newtheorem{lemma}[theorem]{Lemma}

\newtheorem{proposition}{Proposition}

\theoremstyle{definition}
\newtheorem{definition}{Definition}[section]

\newtheorem{claim}{Claim}
\newtheorem{remark}{Remark}

\newtheorem{comment}{Comment}
\newenvironment{proofsketch}{\QuadSpace\par\noindent{\bf Proof sketch}:}{\endproof}
\newcommand{\QuadSpace}{\vspace{0.25\baselineskip}}

\def\RL0{{\mbox{R(lin)}}}
\def\RZ0{{\mbox{R$^0$(lin)}}}
\def\RC0{R(lin) with constant coefficients}

\def\Tse0{{\mbox{$\neg$\textsc{Tseitin}$_{G,p}$}}}

\usepackage[dvipdfm]{color}
\definecolor{bluetxt}{rgb}{0,0,.5}
\definecolor{myred}{rgb}{0.6,0.0,0.1}
\definecolor{greentxt}{rgb}{0,.5,0}

\renewcommand{\vec}{\overline}

	\newcommand{\la}{\langle}
	\newcommand{\ra}{\rangle}
	\newcommand{\Lam}{\mathrm{\Lambda}}
	\newcommand{\BB}{\set{0,1}}
		\newcommand{\circvee}{\hspace{-.1em}\bigcirc\hspace{-1.03em}{\vee}\hspace{.15em}}
\newcommand{\norm}[1]{||#1||}

 \title[Complexity of Propositional Proofs under a Promise]{Complexity of Propositional Proofs\\ under a Promise}
\thanks{This
work was carried out in partial fulfillment of the requirements
for the Ph.D.\ degree of the second author and was supported
in part by the Israel Science Foundation (grant no.~250/05).}
\author{Nachum Dershowitz}
\address{School of Computer
Science, Tel Aviv University, Tel Aviv 69978, Israel}
\email{nachumd@tau.ac.il}
\author{Iddo Tzameret}
\address{School of Computer
Science, Tel Aviv University, Tel Aviv 69978, Israel}
\email{tzameret@tau.ac.il}
\begin{document}

\begin{abstract}
We study -- within the framework of propositional proof complexity -- the
problem of certifying unsatisfiability of CNF formulas under the promise that
any satisfiable formula has many satisfying assignments, where ``many'' stands
for an explicitly specified function $\Lam$ in the number of variables $n$. To
this end, we develop propositional proof systems under different measures of
promises (that is, different $\Lam$) as extensions of resolution. This is done
by augmenting resolution with axioms that, roughly, can eliminate sets of truth
assignments defined by Boolean circuits. We then investigate the complexity of
such systems, obtaining an exponential separation in the average-case between
resolution under different size promises:
\begin{enumerate}
\item Resolution has polynomial-size refutations for all unsatisfiable 3CNF formulas when the
promise is $\eps\!\cd\!2^n$, for any constant $0<\eps<1$.
\item There are no sub-exponential size resolution refutations for random 3CNF
formulas, when the promise is $2^{\delta n}$ (and the number of clauses is
$o(n^{3/2})$), for any constant $0<\delta<1$.
\end{enumerate}
\end{abstract}
\keywords{Propositional proof complexity, Resolution, Random 3CNF, Promise
problems}

\subjclass[2000]{ 03F20, 68Q17, 68Q15}
\maketitle

\begin{quote}\small\raggedleft\it
``Goods Satisfactory or Money Refunded'' \\\rm ---The Eaton Promise
\end{quote}

\section{Introduction}\label{secIntro}
Demonstrating unsatisfiability of propositional formulas is a
fundamental problem in both logic and complexity theory,
as well as in hardware and software validation.
Any standard sound and complete propositional proof system has the
ability to separate the set of unsatisfiable formulas in conjunctive normal
form (CNF) from the set of CNF formulas having at least one satisfying
assignment, in the sense that every unsatisfiable CNF has a refutation in the
system, while no satisfiable CNF has one. Our goal is to develop and study, within
the framework of propositional proof complexity, systems that are ``sound
and complete'' in a relaxed sense: they can separate the set of unsatisfiable
CNF formulas from the set of CNF formulas having \emph{sufficiently many}
satisfying assignments (where the term ``sufficiently many'' stands for an
explicitly given function of the number of variables in the CNF). We  call
such proof systems \emph{promise refutation systems}, as they are complete and
sound for the set of CNF formulas promised to be either unsatisfiable or to
have many satisfying assignments.

As the proof systems we develop here are intended for proving
\emph{unsatisfiability} of CNF formulas (in other words, to \emph{refute} them,
which is the same as validating their negation), throughout this paper we work
solely with \emph{refutation} systems, and speak about ``refutations'' and
``proofs'' interchangeably, always intending refutations, unless otherwise stated.
In particular, we work with refutation systems that extend the widely
studied resolution refutation system.

Our first task is to introduce a natural model for promise propositional
refutation systems. This is accomplished by augmenting standard resolution
(or any other propositional proof system extending
resolution) with an additional collection of axioms, the \emph{promise
axioms}.  Each refutation in a promise refutation system can make use of at
most one promise axiom. The promise axioms are meant to capture the idea that
we can ignore or ``discard'' a certain number of truth assignments from the
space of all truth assignments, and still be able to certify (due to the
promise) whether or not the given CNF is unsatisfiable. The number of
assignments that a promise axiom is allowed to discard depends on the promise
we are given, and, specifically, it needs to be less than the number of
assignments promised to satisfy a given CNF (unless it is unsatisfiable).

Assuming we have a promise that a satisfiable CNF has more than $\Lam$
satisfying assignments, we can discard up to $\Lam$ assignments. We
refer to $\Lam$ as the \emph{promise}. This way, the refutation system is
guaranteed not to contain refutations of CNF formulas having more than
$\Lam$ satisfying assignments, as even after discarding (at most $\Lam$)
assignments, we still have at least one satisfying assignment left.  On the
other hand, any unsatisfiable CNF formula has a refutation in the system, as
resolution already  has a refutation of it.

We now explain (somewhat
informally) what it means to ``discard'' assignments and how promise axioms
formulate the notion of discarding the \emph{correct number} of truth
assignments.
Essentially, we say that a truth assignment $\vec a$ is \emph{discarded} by some
Boolean formula if $\vec a$ falsifies the formula. More formally, let
$X:=\set{x_1,...,x_n}$ be the set of underlying variables of a given CNF,
called the \emph{original variables}. Let $A$ be some CNF formula in the variables $X$,
and assume that $A$ also contains variables not from $X$, called
\emph{extension variables}. Let $\vec{a}\in\BB^n$ be a truth assignment for the
$X$ variables, and assume that there is no extension of $\vec{a}$ (assigning values to the
extension variables) that satisfies $A$. Thus, any assignment satisfying $A$
must  also satisfy $X\not\equiv\vec{a}$ (that is, $A\models
X\not\equiv\vec{a}$), and so any (implicationally) complete proof system can
prove $X\not\equiv\vec{a}~$ from $A$, or, in the case of a refutation system,
can refute $X\equiv\vec{a}$, given $A$. In this case, we say that the
assignment $\vec{a}$ is \emph{discarded by $A$}.

The promise axioms we present enjoy two main properties:
\begin{enumerate}
  \item They discard assignments from the space of possible assignments to the variables $X$.
  \item They express the fact that not too many assignments to the variables $X$ are
   being discarded (in a manner made precise).\label{itProp}
\end{enumerate}

The first property is achieved as follows: Let $C$ be any Boolean circuit with
$n$ output bits. Then we can formulate a CNF formula $A$ (using
extension variables) expressing the statement that the output of $C$ is (equal to)
the vector of variables
$X$. This enables $A$ to \emph{discard every
truth assignment to the variables of $X$ that is outside the image of the Boolean map
defined by $C$},
because, if an assignment $\vec{a}$ to $X$ is not in
the image of $C$, then no extension of $\vec{a}$ can satisfy $A$---assuming the
formulation of $A$ is correct. (For technical reasons, the actual definition is a bit different than
what is described here; see Section~\ref{secMod}.)

The second property is achieved as follows: Assume we can make explicit the statement
that the \emph{domain} of the map defined by the Boolean circuit $C$ above is
of size at least $2^n-\Lambda$. (See Section~\ref{secMod} for more
details.) Then,  for the second property to hold, it is
sufficient that the axiom formulates the statement that the circuit $C$ defines
an \emph{injective} map (and thus the image of the map contains enough truth
assignments), which can be done quite naturally.

Given a certain promise and its associated promise axiom, we call a
refutation of resolution, augmented with the promise axiom, a \emph{resolution
refutation under the (given) promise}.

Our second task, besides introducing the model of promise
refutation systems, is to investigate the basic properties of this model and in
particular to determine its average-case proof complexity with respect to
different size of promises (see below for a summary of our findings in this
respect).

\subsection{Background and Motivation}\label{secBG}

In propositional proof complexity theory, it is standard to consider an
\emph{abstract }or \emph{formal } propositional proof system (usually called
\emph{a Cook-Reckhow proof system}, following~\cite{CR79}) as a polynomial-time
algorithm $A$ that receives a Boolean formula $F$ (usually in CNF) and a string
$\pi$ over some finite alphabet (``the (proposed) refutation'' of $F$), such
that there exists a $\pi$ with $A(F,\pi)=1$ if and only if $F$ is
unsatisfiable. (A string $\pi$ for which $A(F,\pi)=1$ is also called a
\emph{witness} for the unsatisfiability of $F$.) Equipped with this abstract
definition of propositional proof systems, showing that \emph{for every}
abstract proof system there exists some family of formulas $F$ for which there
is no polynomially-bounded family of proofs $\pi$ of $F\,$ is equivalent to
showing $\NP\neq\coNP$.

For this reason (among others), it is customary in proof complexity theory to
concentrate on specific (sometimes provably weaker) proof systems for which
proofs have a simple structure. This makes the complexity analysis of such
proof systems simpler. Prominent examples of such systems are Frege systems and
weaker subsystems of Frege, the most notable being the resolution refutation
system~\cite{Rob65}, which also plays an important r\^{o}le in many automated
theorem provers. In accordance with this, we shall be interested in the present
paper not with abstract proof systems (that is, not with finding general
witnesses for unsatisfiability, possibly under a promise), but rather with
specific and more structured proof systems, and specifically with refutation
systems built up as extensions of resolution.

%

A natural relaxation of the problem of unsatisfiability certification is to
require that, if a CNF is satisfiable, then it actually has many satisfying
assignments. As mentioned above, we call the specific number of assignments (as
a function of the number of variables $n$) required to satisfy a satisfiable
CNF formula, the ``promise''. Accordingly, one can define an \emph{abstract
promise proof system} in an analogous manner to the definition of an abstract
proof system.
It is thus natural to ask whether giving such a promise can help in obtaining
shorter proofs of unsatisfiability.

In the case of a \emph{big} promise, that is, a constant fraction of the space
of all truth assignments ($\Lambda=\eps\cd 2^n$, for a constant $0<\eps<1$),
there is already a \emph{deterministic polynomial-time algorithm} for any fixed
natural number $k$ that certifies the unsatisfiability of all unsatisfiable
$k$CNF formulas under the promise: The algorithm receives a $k$CNF that is
either unsatisfiable or has more than $\Lambda$ satisfying assignments and
answers whether the formula is unsatisfiable (in case the formula is
satisfiable the algorithm provides a satisfying assignment). See
\cite{Hir98,Tre04} for such efficient algorithms.\footnote{In
the case the promise is $\Lambda=2^n/poly(n)$, the algorithm in~\cite{Hir98}
also gives a deterministic sub-exponential time procedure for unsatisfiability
certification of $k$CNF formulas (for a constant $k$).} This trivially implies
the existence of polynomial-size witnesses for any unsatisfiable $k$CNF under
the promise $\eps\cd 2^n$. But does resolution already admit such short
witnesses of unsatisfiability (that is, resolution refutations) under a big
promise? We show that the answer is positive (for all unsatisfiable 3CNF
formulas).

In the case of a \emph{smaller} promise, by which we mean $\Lambda=2^{\delta
n}$ for a constant $0<\delta<1$, it is possible to efficiently transform any
CNF over $n$ variables to a new CNF with $n'=\lceil n/(1-\delta)\rceil$
variables, such that the original CNF is satisfiable if and only if the new CNF
has at least $2^{\delta n'}$ satisfying assignments.\footnote{
This can be achieved simply by
adding new ($n'-n$) ``dummy variables''. For instance,
by adding the clauses of a tautological CNF in these dummy variables to
the original CNF. This way, if the original CNF has at least one satisfying
assignment then the new CNF has at least $2^{n'-n} \ge 2^{\delta n'}$
satisfying assignments.}
 Thus, the \emph{ worst-case} complexity of certifying CNF
unsatisfiability under such a promise is polynomially equivalent to the
worst-case complexity of certifying CNF unsatisfiability without a promise.
However, it is still possible that a promise of $2^{\delta n}$ might give some
advantage (that is, a super-polynomial speedup over refutations without a
promise) in certifying the unsatisfiability of certain (but not all) CNF
formulas; for instance, in the
\emph{average-case}.\footnote{Note that if we add dummy
variables to a 3CNF then we obtain an ``atypical instance'' of a 3CNF. Thus,
assuming we have polynomial-size witnesses of unsatisfiability of 3CNF formulas
under a small promise in the average-case (that is, the ``typical case''), the
reduction alone (that is, adding dummy variables) does \emph{not} automatically
yield polynomial-size witnesses for 3CNF formulas in the average-case without a
promise as well.}

Feige, Kim, and Ofek~\cite{FKO06} have shown that when the number of
clauses is $\mathrm\Omega(n^{7/5})$ there
exist polynomial-size witnesses to the unsatisfiability of 3CNF
formulas in the \emph{average-case}.
On the other hand, Beame,  Karp, Pitassi, and Saks~\cite{BKPS02} and Ben-Sasson and
Wigderson~\cite{BSW99} showed that resolution does
not provide sub-exponential refutations for 3CNF formulas in the average-case
when the number of clauses is at most $n^{(3/2)-\epsilon}$, for any constant
$0<\epsilon<1/2$.\footnote{Beame \emph{et al.}~\cite{BKPS02} showed such a lower
bound for $n^{(5/4)-\epsilon}$ number of clauses (for any constant
$0<\epsilon<1/4$). Ben-Sasson and
Wigderson~\cite{BSW99} introduced the size-width tradeoff that enables
one to prove an exponential lower bound for random 3CNF formulas with
$n^{(3/2)-\epsilon}$ number of clauses (for any constant $0<\epsilon<1/2$), but
the actual proof for this specific clause-number appears in~\cite{BS-Phd}.}
This shows that general witnessing of 3CNF unsatisfiability is strictly
stronger than resolution refutations. But is it possible that, under a promise
of $2^{\delta n}$, resolution can do better in the average-case? We show that
the answer is negative.

There are two main motivations for studying propositional proofs under a
given promise and their complexity. The first is to answer the natural question
whether CNF unsatisfiability certification enjoys any advantage given a certain
promise.
As already mentioned, the answer is positive when the promise is a constant
fraction of all the truth assignments, and our results imply that this
phenomenon already occurs for resolution. For a small promise of $2^{\delta
n}$, we can show that, at least in the case of resolution refutations of most
3CNF formulas (of certain clause-to-variable density), the answer is negative.
In fact, we can show that the answer stays negative even when the promise is
bigger than $2^{\delta n}$, and specifically when $\Lam=2^n/2^{n^\xi}$ for some
constant $0<\xi<1$.
Overall, our results establish the first unsatisfiability certification model
in which a promise of a certain given size is known to help
(that is, allow more efficient certifications) in the average-case,
while promises of smaller sizes do not help.

The second motivation is more intrinsic to proof complexity theory: It is a
general goal to develop natural frameworks for propositional proofs that are
not sound in the strict sense, but rather possess an approximate notion of
soundness (like showing that certain ``approximations'' give speed-ups).
 For this purpose, the proof systems we propose formalize%
---in a natural way---the notion of separating
unsatisfiable CNF formulas from those that have many satisfying
assignments. The promise axioms we present also allow for a natural way of
controlling the size of the promise, which in addition leads to an exponential
separation between different size promises.


This paper introduces the concept of propositional proofs under a promise,
analyzes the proof complexity of these proof systems with respect to different
promise sizes, giving a
separation between promises of different sizes,
and also illustrates several new facts about the widely studied
resolution proof system.

\subsection{Results}
We show that resolution refutations are already enough
to efficiently separate unsatisfiable 3CNF formulas from those 3CNF formulas
with an arbitrarily small constant fraction of satisfying assignments. In
particular, in Section~\ref{secBigPro}, we show the following:

\subsubsection*{Main Result 1:} Let $0<\eps<1$ be any constant and let
$\Lambda=\eps\!\cd\!2^n$ be the given promise. Then \emph{every} unsatisfiable
3CNF with $n$ variables has a polynomial-size (in $n$) resolution refutation
under the promise $\Lambda$.

The proof of this resembles a deterministic algorithm of Trevisan~\cite{Tre04}
for approximating the number of satisfying assignments of $k$CNF
formulas. 

In contrast to the case of a big promise, the results show that, at least for
resolution, a small promise of $\Lambda=2^{\delta n}$ (for any constant
$0<\delta<1$) does not give any advantage over standard resolution (that is,
resolution without the promise axioms) in most cases (that is,
in the average-case).
Specifically, in Section~\ref{secSmlPro} we show the following:

\subsubsection*{Main Result 2:} Let $0<\delta<1$ be any constant and let
$\Lambda=2^{\delta n}$ be the given promise. Then, there is an exponential lower
bound on the size of resolution refutations of random 3CNF formulas under the
promise $\Lambda$, when the number of clauses is $\SmallO(n^{3/2})$.

This lower bound actually applies to a more general model of promise proofs.
It remains valid even if we allow (somehow) the promise proofs to discard
\emph{arbitrarily chosen} sets of truth assignments (of size $\Lam=2^{\delta n}$),
and not necessarily those sets that are definable by (small) Boolean
circuits. In fact, the lower bound applies even to a bigger promise of
$\Lam=2^{n-n^\xi}$, for some constant $0<\xi<1$.

The proof strategy for this lower bound follows that of Ben-Sasson and
Wigderson~\cite{BSW99} (the
\emph{size-width tradeoff} approach), and so the rate of the lower bound
matches the one in that paper. The main novel observation is that under the
appropriate modifications this strategy also works when one restricts the set
of all truth assignments to a smaller set (that is, from $2^n$ down to
$2^n-2^{\delta n}$ for a constant $0<\delta<1$, and in fact down to
$2^n-2^n/2^{n^\xi}$, for some constant $0<\xi<1$).

It is important to note that these two main results show that the decision
to discard sets of truth assignments defined by \emph{Boolean circuits} does
not affect the results in any way, and thus should not be regarded as a
restriction of the model of promise refutations (at least not for resolution).
To see this, note that we could allow a promise refutation to discard
\emph{arbitrarily chosen} sets of truth assignments (of the appropriate size
determined by the given promise), that is, sets of truth assignments that are
not necessarily definable by (small) Boolean circuits. However, although this
modification strengthens the model, it is not really necessary for the
\emph{upper bound} in Main Result 1, as this upper bound is already valid when
one discards sets of truth assignments by (small) Boolean circuits.
On the other hand, as mentioned above, the \emph{lower bound }in Main Result 2 is
already valid when one allows a promise refutation to discard any
\emph{arbitrarily chosen} set of truth assignments (of the appropriate size).

The exact model of promise propositional proof systems is developed in Section
\ref{secMod}. It is preceded, in the next section, by preliminaries and terminological conventions.

\section{Preliminaries} \label{secPre}

\subsection{Notations}

For natural number $m$, we use $[m]$ to denote the set $\set{1,\ldots, m}$ of naturals.

Let $A,B$ be two propositional formulas. We write $\,A \equiv B\,$ as an
abbreviation for $(A\to B)\wedge(B\to A)$. The notation $A\not\equiv B$
abbreviates $\neg(A \equiv B$). We say that \emph{$A$ semantically implies
$B$}, denoted by $A\models B$, iff every satisfying assignment to $A$ also
satisfies $B$.


A CNF formula over the variables $x_1, \ldots, x_n$ is defined as follows: A
\emph{literal} is a variable $x_i$ or its negation $\neg x_i$. A \emph{clause }
is a disjunction of literals. We treat a clause as a set of literals, that is,
we delete multiple occurrences of the same literal in a clause. A CNF formula
is a conjunction of clauses (sometimes treated also as a \emph{set} of clauses,
where the conjunction between these clauses is implicit). A $k$CNF formula is a
CNF with all clauses containing $k$ literals each.

The \emph{width} of a clause $D$ is the number of literals in it, denoted
$|D|$. The \emph{size} of a CNF formula $K$ is the total number of clauses in
it, denoted $|K|$. The \emph{width of a CNF formula} $K$ is the maximum
width of a clause in $K$.

We denote by $K'\subseteq K$ that $K'$ is a sub-collection of clauses from $K$.

\subsection{Resolution Refutation Systems}

Resolution is a complete and sound proof system for unsatisfiable CNF formulas.

Let $C$ and $D$ be two clauses containing neither $x_i$ nor $\neg x_i$. The
\emph{resolution rule} allows one to derive $C\vee D$ from $C\vee x_i$ and
$D\vee \neg x_i$. The clause $C\vee D$ is called the \emph{resolvent} of the
clauses $C\vee x_i$ and $D\vee \neg x_i$ on the variable $x_i$, and we also say
that $C\vee x_i$ and $D\vee \neg x_i$ were \emph{resolved over $x_i$}.

The \emph{weakening rule} allows one to derive the clause $C\vee D$ from the clause
$C$, for any two clauses $C,D$. 

\begin{definition}[{Resolution}]\label{defnRes}
A \emph{resolution proof of the clause $D$ from a CNF formula $K$} is a
sequence of clauses $D_1,D_2,\ldots,D_\ell\,$, such that: (1) each clause $D_j$
is either a clause of $K$ or a resolvent of two previous clauses in the
sequence or derived by the weakening rule from a previous clause in the
sequence; (2) the last clause $D_\ell=D$. The size of a resolution proof is the
total number of clauses in it. The width of a resolution proof is the maximal
width of a clause in it. A \emph{resolution refutation} of a CNF formula $K$ is
a resolution proof of the empty clause $\Box$ from $K$. (The empty clause stands
for \textsc{false}; that is, the empty clause has no satisfying assignments.)
\end{definition}

Let $K$ be an unsatisfiable CNF formula. The \emph{resolution refutation size
of $K$} is the minimal size of a resolution refutation of $K$ and is denoted
$S(K)$. Similarly, the \emph{resolution refutation width of $K$} is the minimal
width of a resolution refutation of $K$ and is denoted $w(K)$. If $K$ has a
polynomial-size resolution refutation we say that resolution can
\emph{efficiently certify }the unsatisfiability of $K$. Similarly, if the
clause $D$ has a polynomial-size resolution proof from $K$ we say that
\emph{$D$ is efficiently provable from $K$}.

\subsection{Size-Width Tradeoffs}

We recall now the approach for proving size lower bounds on
resolution refutations developed by Ben-Sasson and Wigderson
\cite{BSW99}. The basic idea is that a
lower bound on the resolution refutation width of a CNF formula $K$ implies a
lower bound on the resolution refutation size of $K$:

\begin{theorem}[\cite{BSW99}]\label{thmBSW} Let $K$ be a CNF formula of width $r$, then
$$ S(K)=\exp\left(\Omega\left(\frac{(w(K)-r)^2}{n}\right)\right).$$
\end{theorem}

\subsection{Boolean Circuit Encoding}

The promise axioms we introduce use Boolean circuits to define the set of
assignments to be discarded (see Section~\ref{secMod}).
Therefore, as resolution operates only with clauses,
we need to encode Boolean circuits as collections of clauses (CNF formulas).
We assume that all Boolean circuits use only three gates: $\vee,\wedge,\neg$
(though this is not necessary) where $\vee$ (denoting \textsc{or}) and $\wedge$
(denoting \textsc{and}) have fan-in $2$ and $\neg$ (denoting \textsc{not}) has
fan-in $1$. Let $C$ be a Boolean circuit with $m$ input bits and $n$ output
bits. Let $\vec{W}=\set{w_1,\ldots,w_m}$ be the $m$ input variables of $C$ and
let $X$ denote the $n$ variables $\set{x_1,\ldots, x_n}$. We consider the $n$
output bits of $C$ as the outputs of $n$ distinct circuits
$C_1(\vec{W}),\ldots,C_n(\vec{W})$ in the $\vec{W}$ variables, and we write
$C(\vec W)\equiv X$ to mean that $X$ equals the output of $C(\vec W)$ (that is,
$C_1(\vec W)\equiv x_1\wedge \cdots \wedge C_n(\vec W)\equiv x_n$). This
notation can be  extended  in a similar manner to $C(\vec{W_1})\equiv
C'(\vec{W_2})$ and $C(\vec{W_1})\not\equiv C'(\vec{W_2})$.


By Cook's Theorem, there exists a CNF formula $F$ (in both the $\vec{W}$
variables and new extension variables) that \emph{encodes the circuit $C$}.
This means that there are $n$ new  extension variables (among other extension
variables) $y_1,\ldots, y_n$ in $F$ such that for all assignments $\vec{a}$:\,
$F(\vec{a})=1$ iff
$C(w_1(\vec{a}),\ldots,w_m(\vec{a}))=y_1(\vec{a})\circ\cdots\circ y_n(\vec{a})$,
where we denote by $w_i(\vec{a})$ the truth value of $w_i$ under the
assignment $\vec{a}$ and by $\circ$ the concatenation of Boolean bits. In
other words, $F$ expresses the fact that $y_1,\ldots,y_n$ are the output bits
of $C$. If $C$ is of size $s$ (that is, the number of Boolean gates in $C$ is
$s$), then the size of $F$ is $O(s\cd\log(s))$. Therefore, if $C$ is of size
polynomial in $n$ then $F$ is also of polynomial-size in $n$. We denote by
$\|C(\vec{W})\|$ the CNF formula $F$ that encodes $C(\vec{W})$.

For most purposes, we will not need an explicit description of how the encoding
of Boolean circuits as CNF formulas is done through $\|C(\vec{W})\|$.
Nevertheless, in Section~\ref{secBigPro} we need to ensure that resolution can
efficiently prove several basic facts about the encoded circuits. For this
reason, and for the sake of concreteness of the promise axioms (Definitions
\ref{defAx} and~\ref{defAx2})  we provide the precise definition of the
encoding in the Appendix (Section~\ref{secEnc}), in addition to proving some of
the encoding's basic (proof theoretical) properties. The interested reader can
look at the Appendix for any missing details, but anyone willing to accept the
existence of an efficient CNF encoding of Boolean circuits that is also
intensional for resolution (in the sense that resolution can efficiently prove
basic properties of the encoded circuits) can skip Section~\ref{secEnc} without
risk.

\section{Promise Proof Systems}\label{secMod}

In this section we define precisely the model of refutations under a promise.
As discussed in the introduction, we work with
the resolution refutation system as our underlying system and augment it with
a new set of axioms that we call the \emph{promise axioms}. We call this proof
system \emph{promise resolution}. The promise axioms are meant to express the
fact that we can discard a certain number of truth assignments from the space
of all truth assignments and still be able to certify (due to the promise)
whether the input CNF is unsatisfiable or not. Each promise resolution
refutation can use at most one promise axiom.

From now on, throughout the paper, we shall assume that the underlying
variables of the CNF formulas that are meant to be refuted are taken from the
set $X:=\set{x_1,\ldots,x_n}$. The $X$ variables are called the \emph{original
variables}. Any other variable that appears in a (promise resolution)
refutation is called an \emph{extension variable}.

\begin{definition}[{CNF formulas under a promise}]\label{defCNFpro}
Let $\Lambda$ be a fixed function
in $n$ (the number of $X$ variables) such that $0\leq \Lambda(n) \leq 2^n$. The function
$\Lambda$ is called the \emph{promise}. The set of \emph{CNF formulas under the promise
$\Lambda$} consists of all CNF formulas in the $X$ variables that are either unsatisfiable or
have more then $\Lambda(n)$ satisfying assignments (for $n=|X|$).
\end{definition}

The refutation systems we build are sound and complete for the set of CNF
formulas under a (given) promise. That is, every unsatisfiable CNF formula has
a refutation in the system (this corresponds to completeness), while no CNF
having $n$ variables and more than $\Lambda(n)$ satisfying assignments has a
refutation in it (this corresponds to soundness under the promise).
Soundness (under the promise) is achieved by requiring that resolution should
\emph{prove the fact that we discard the right number of assignments} (see
Section~\ref{secAx} for details).

\begin{definition}[{Assignment discarding}]
Let $A$ be a CNF in the $X$ variables that can contain (but does not necessarily
contain) extension variables (that is, variables not from $X$). We say that an
assignment to the $X$ variables $\vec{a}$ is \emph{discarded} by $A$ if there
is no extension of $\vec{a}$ (to the extension variables in $A$) that satisfies
$A$.
\end{definition}
\noindent (See Section~\ref{secIntro} for more regarding assignment discarding.)

\subsection{Promise Axioms}\label{secAx}

\subsubsection{Big promise}
We first concentrate on a promise of a \emph{constant fraction of assignments}
(for a smaller promise the axiom is similar; see below).

Let the promise (see Definition~\ref{defCNFpro}) be $\Lambda=\eps\cd 2^n$, for
a constant $0<\eps<1$ (we fix this $\Lambda$ throughout this subsection), and
let $r=\lceil\log(1/\eps)\rceil$ and $t=2^r-1$. Let $C$ be a sequence of
Boolean circuits $C:=(C^{(1)},\ldots,C^{(t)})$. Assume that each $C^{(i)}$ has
$n-r$ input bits and $n$ output bits and computes the Boolean map
$f_i:\BB^{n-r}\to\BB^n$. Assume further that the $f_i$'s are all injective maps
and that the images of all these maps are pairwise disjoint. Denote by
$\Im(f_i)$ the image of the map $f_i$.
For simplicity, we call the union $\cup_{i=1}^{t}{\Im}(f_i)$
\emph{the image of $C$} and denote it by $\Im(C)$. By the definition of $r$, we
have $2^{n-r}\leq \eps\cd2^n$, and by the injectivity and pairwise disjointness
of the images of the $f_i$'s we have:
\begin{equation}\label{eqImg}
|{\Im}(C)| = t\cd 2^{n-r} = (2^r-1)\cd 2^{n-r} = 2^n-2^{n-r} \geq 2^n-\Lambda\,.
\end{equation}
Therefore, \emph{we can treat ${\Im}(C)$ as the set of all possible truth
assignments for the original variables $X$, without losing soundness}: If $K$
is unsatisfiable then there is no assignment in $\Im(C)$ that satisfies $K$; and
if $K$ is satisfiable then according to the promise it has more than $\Lambda$
satisfying assignments, which means that there is at least one assignment in
$\Im(C)$ that satisfies $K$. This idea is formulated as a propositional formula
as follows:

\begin{definition}[{Promise Axiom for $\Lambda=\eps\cd 2^n$}]\label{defAx}
Let the promise be $\Lambda=\eps\cd 2^n$, for a constant $0<\eps<1$, and let
$r=\lceil\log(1/\eps)\rceil$ and $t=2^r-1$. Let $C$ be a sequence of Boolean
circuits $C:=(C^{(1)},\ldots,C^{(t)})$. Assume that each $C^{(i)}$ has $n-r$
input bits and $n$ output bits and let $\vec{W_1}$ and $\vec{W_2}$ be two
disjoint sets of $\,n-r\,$ extension variables each. The promise axiom
\emph{PRM}$_{C,\Lambda}$ is the CNF encoding (via the encoding defined in
Section~\ref{secEnc}) of the following Boolean formula:
\begin{equation}
\begin{array}{r}\label{eq01}
\left( \bigwedge\limits_{i=1}^{t} \left(C^{(i)}(\vec{W_1})\equiv
C^{(i)}(\vec{W_2}) \to \vec{W_1} \equiv \vec{W_2}\right) \wedge
\bigwedge\limits_{1\le i< j\le t} C^{(i)}(\vec{W_1})\not\equiv
C^{(j)}(\vec{W_2})\right)
\nonumber \\
\qquad\qquad\qquad\qquad\qquad \longrightarrow\bigvee\limits_{i=1}^{t} C^{(i)}(\vec{W_1})
\equiv X.
\end{array}
\end{equation}
\end{definition}

The promise axiom PRM$_{C,\Lambda}$ expresses the fact that if each circuit in
$C$ computes an injective map (this is formulated as
$\wedge_{i=1}^{t}(C^{(i)}(\vec{W_1})\equiv C^{(i)}(\vec{W_2})\to \vec{W_1}
\equiv \vec{W_2})$), and if the images of the maps computed by each pair of
circuits in $C$ are disjoint (this is formulated as $\wedge_{1\le i< j\le t}
C^{(i)}(\vec{W_1})\not\equiv C^{(j)}(\vec{W_2})$), then we can assume that the
assignments to the original variables $X$ are taken from the image of $C$ (this
is formulated as $\vee_{i=1}^{t} C^{(i)}(\vec{W_1})\equiv X$). The fact that
the image of $C$ is of size at least $2^n-\Lambda$ is expressed (due to
Equation (\ref{eqImg})) by the number of input bits (that is, $n-r$) of each
circuit in $C$ and the number of circuits in $C$ (that is, $t$). Also note that
the promise axiom is of polynomial-size as long as the circuits in $C$ are
(since $1/\eps$\, is a constant).

The following claim shows that the promise axioms are sound with respect to the
promise $\Lambda$,  in the sense that they do not discard too many truth
assignments:

\begin{claim}\label{claCor}
The promise axiom \emph{PRM}$_{C,\Lambda}$ discards at most $\Lambda$ truth
assignments. That is, there are at most $\Lambda$ distinct assignments
$\vec{a}$ to the $X$ variables such that \emph{PRM}$_{C,\Lambda} \models
X\not\equiv \vec{a}$.
\end{claim}
\begin{proof}
Assume that some Boolean map computed by some circuit in $C$ is not injective.
Then any assignment to the $X$ variables has an extension $\rho$ (to the
extension variables in the promise axiom) that falsifies the premise of the
main implication in PRM$_{C,\Lambda}$ and thus $\rho$ satisfies
PRM$_{C,\Lambda}$. Therefore no assignments to $X$ are discarded.

Similarly, assume that the images of some pair of maps computed by two circuits
in $C$ are not disjoint. Then, again, any assignment to the $X$ variables has
an extension that satisfies PRM$_{C,\Lambda}$, and so no assignments to $X$ are
discarded.

Assume that all the Boolean maps computed by circuits in $C$ \emph{are}
injective and have pairwise disjoint images. Then every assignment satisfies
the premise of the main implication in the promise axiom PRM$_{C,\Lambda}$.
Therefore, it suffices to show that the consequence of the main implication of
the axiom (that is,  $\,\vee_{i=1}^{t} C^{(i)}(\vec{W_1})\equiv X$\,) discards
at most $\Lambda$ assignments to the $X$ variables. By definition (of the
encoding of the circuits) for all assignments $\vec{a}$ to the $X$ variables
that are in $\Im(C)$ there is an extension of $\vec{a}$ that satisfies
$\vee_{i=1}^{t} C^{(i)}(\vec{W_1})\equiv X$. Now, all the circuits $C^{(i)}$
compute injective maps with pairwise disjoint images, and thus by Equation
(\ref{eqImg}) there are at least $2^n-\Lambda$ distinct elements (that is,
assignments) in $\Im(C)$. Hence, at least $2^n-\Lambda$ assignments to the $X$
variables are not discarded.
\end{proof}

\subsubsection{Smaller promise}
We shall also need to formulate promise axioms for promises smaller than
$\eps\cd 2^n$. Specifically, we shall work with a promise of $\Lambda=2^{\delta
n}$ for a  constant $0<\delta<1$ (we fix this $\Lambda$ throughout this
subsection). For such a promise, the promise axiom is similar to Definition
\ref{defAx}, except that the number of input bits of each circuit in $C$ needs
to be modified accordingly. (We shall use the same terminology as that used
above for the Big Promise.)

\begin{definition}[{Promise Axiom for $\Lambda=2^{\delta n}$}]\label{defAx2}
Let the promise be $\Lambda=2^{\delta n}$, for a  constant $1<\delta<1$, and
let $t=\left\lceil {(1 - \delta )n} \right\rceil$. Let $C$ be a sequence of
Boolean circuits $C:=(C^{(1)},\ldots,C^{(t)})$. Assume that for each $1\leq
i\leq t$ the circuit $C^{(i)}$ has $n-i$ input bits and $n$ output bits. Let
$\vec{W_1},\ldots,\vec{W_t}$ and $\vec{W'_1},\ldots,\vec{W'_t}$ be $2t$
disjoint sets of extension variables\footnote{We have not been very economical in adding extension variables
here; but this is not essential.}, where for all $1\leq i\leq
t$, $W_i, W'_i$ consist of $\,n-i\,$ variables each. The promise axiom
\emph{PRM}$_{C,\Lambda}$ is the CNF encoding (via the encoding defined in
Section~\ref{secEnc}) of the following Boolean formula:
\begin{equation}
\begin{array}{r}\label{eq03}
\left( \bigwedge\limits_{i=1}^{t} \left(C^{(i)}(\vec{W_i})\equiv
C^{(i)}(\vec{W'_i}) \to \vec{W_i} \equiv \vec{W'_i}\right) \wedge
\bigwedge\limits_{1\le i< j\le t} C^{(i)}(\vec{W_i})\not\equiv
C^{(j)}(\vec{W_j})\right)
\nonumber \\
\qquad\qquad\qquad\qquad\qquad \longrightarrow\bigvee\limits_{i=1}^{t}
C^{(i)}(\vec{W_i})\equiv X.
\end{array}
\end{equation}
\end{definition}
Note that the promise axiom is of
polynomial size as long as the circuits in $C$ are (since $t\leq n$).

Also note that the proof of Claim~\ref{claCor} did not use the parameters $r$
and $t$ (which determine the number of input bits in the circuits in $C$ and
the number of circuits in $C$, respectively) but only the size $|$$\Im(C)|$.
Thus, the same claim holds also for the promise axiom in Definition
\ref{defAx2}, which means that this promise axiom discards at most
$2^n-|$$\Im(C)|$ truth assignments, for some sequence of circuits in $C$ that
compute injective maps with pairwise disjoint images. Therefore, we need to
verify that $|{\Im}(C)|\ge2^n-\Lambda$, for all $C$ that consists of
circuits computing injective maps with pairwise disjoint images.

Notice that for all $1\leq i\leq t$ the circuit $C^{(i)}$ computes a Boolean
map, denoted $f_i$, such that $f_i:\BB^{n-i}\to\BB^n$. Assume that all the
$f_i$'s are injective and that the images of each pair of functions $f_i,f_j$,
for $1\le i\neq j\le t$, are disjoint. Then, we have:
\begin{eqnarray}\label{eqHarm}
|{\Im}(C)| &=&
\left( {\frac{1}{2} + \frac{1}{{2^2 }} + \frac{1}{{2^3 }} +  \cdots  +
\frac{1}{{2^t }}} \right) \cdot 2^n  ~=~ \left(1 - \frac{1}{{2^t }}\right) \cdot
2^n\\\nonumber &=& 2^n - 2^{n - \left\lceil {(1 - \delta )n}
\right\rceil }
~ \ge~ 2^n  - 2^{\delta n}  ~=~ 2^n  - \Lambda \nonumber
\end{eqnarray}
Also note that $|{\Im}(C)|\leq 2^n-2^{\delta n -1}$ and so if the circuit
in $C$ are injective with pairwise disjoint images then PRM$_{C,\Lambda}$
discards \emph{at least} $\,2^{\delta n}/2\,$ truth assignments.

\subsection{Promise Resolution}

\begin{definition}[{Promise resolution}]\label{defProRes}
Let $\Lambda$ be the promise (see Definition~\ref{defCNFpro}) and let $K$ be a
CNF in the $X$ variables. A \emph{promise resolution (under the promise
$\Lambda$) proof of the clause $D$ from a CNF formula $K$} is a sequence of
clauses $D_1,D_2,\ldots,D_\ell\,$ such that:
\begin{enumerate}
\item[(1)] Each clause $D_j$ is either a
clause of $K$ or a clause of a promise axiom \emph{PRM}$_{C,\Lambda}$ (where
\emph{PRM}$_{C,\Lambda}$ is either a big or a smaller promise axiom as defined
in Definitions~\ref{defAx} and~\ref{defAx2} and $C$ is an arbitrary sequence of
circuits with the prescribed input and output number of bits) or a resolvent of
two previous clauses in the sequence; \item[(2)] The sequence contains (the clauses
of) at most one promise axiom; \item[(3)] The last clause $D_\ell=D\,$.
\end{enumerate}

The \emph{size}, \emph{width }and \emph{refutations} of promise resolution is
defined the same as in resolution.
\end{definition}

Note that promise resolution is a Cook-Reckhow proof system (see the first
paragraph in Section~\ref{secBG} for a definition): It is possible to
efficiently verify whether a given CNF is an instance of the promise axiom,
and hence to verify whether a sequence of clauses constitute a legitimate promise
refutation.
This can be done by ``decoding" the CNF that encodes the promise axiom
PRM$_{C,\Lambda}$ and then checking that each circuit in $C$ has the right
number of input and output bits (we discuss this issue in some more detail in
the appendix).

\begin{proposition}
Let $\Lambda$ be the promise (where $\Lambda$ is either $\eps\!\cd\! 2^n$ or
$2^{\delta n}$, for $0<\eps,\delta<1$). Then promise resolution under the
promise $\Lambda$ is a sound and complete proof system for the set of CNF
formulas under the promise $\Lambda$ (see Definition~\ref{defCNFpro}). In other
words, every unsatisfiable CNF has a promise resolution refutation and every
CNF that has more than $\Lambda$ satisfying assignments does not have promise
resolution refutations.
\end{proposition}

\begin{proof}
Completeness stems from completeness of resolution. Soundness under the promise
$\Lambda$ stems from Claim~\ref{claCor} (which, by the notes after Definition
\ref{defAx2}, holds for both the big and the smaller promise axioms).
\end{proof}

\subsection{Discussion}

Let $K$ be an unsatisfiable CNF
formula in $n$ variables, PRM$_{C,\Lambda}$ a promise axiom (where the circuits
in $C$ all compute injective and pairwise disjoint Boolean maps) and let
$S:={\Im}(C)\subseteq\BB^n\,$ (such that $|S|\geq 2^n-\Lambda$). Then, one
can think of a promise resolution refutation of $K$ using the axiom
PRM$_{C,\Lambda}$ as containing two separate parts:
\begin{enumerate}
\item[(i)] a resolution `refutation' of $K$ where the space of truth assignments is
restricted to $S$;
\item[(ii)] a resolution proof that $|S|\geq 2^n-\Lambda$.
\end{enumerate}

Note that if we want to consider promise resolution as having only part (i),
then we can modify (actually, strengthen) the promise axiom into
$\vee_{i=1}^{t} C^{(i)}(\vec{W})\equiv X$. However, this choice means losing
the soundness of the proof system under the promise (that is, the soundness
with respect to CNF formulas under a promise as defined in Definition
\ref{defCNFpro}), since we do not have any guarantee that the circuit $C$
discards at most $\Lambda$ assignments (and so CNF formulas with more than
$\Lambda$ satisfying assignments might have refutations in such a system).

It is possible to use any number of axioms of the form $C^{(i)}(\vec{W})\equiv
X$, as long as resolution can prove both the injectivity of each of the maps
computed by the circuits $C^{(i)}$ introduced and the pairwise disjointness of
these maps (as formulated by a propositional formula similar to the formulation
in the promise axioms), and provided that the circuits $C^{(i)}$ have number of
input bits that induce the right size of domains (that is, that the total size
of their domains is at least $2^n-\Lambda$).

It is also possible to modify the promise axioms to suit any chosen size of
promise $\Lambda$ (possibly, only an approximation of $\Lambda$). This can be
achieved by choosing a sequence of circuits with the appropriate size of domain
(explicitly expressed by the number of input bits in each circuit in the
sequence, and the total number of circuits).

Some comments about the formulation of the promise axioms are in order.

\begin{comment}
Note that we could not use only a single circuit $C$ in the promise axioms
(in contrast to a sequence of circuits), because that way we would not have the
possibility of controlling the size of the domain of $C$ and efficiently verifying
that this size is the correct one inside resolution. To see this, note that if
the number of input variables to $C$ is $n$ (the number of original variables)
and the map computed by $C$ is (provably) injective then $C$ does not discard
any assignment. If, on the other hand, the number of input variables to $C$ is
less than $n$, then $C$ discards at least half the truth assignments, which
might be too many.
\end{comment}

\begin{comment}
Also note that in order to discard assignments we cannot use a seemingly
more natural axiom of the form $C(\vec{W})\not\equiv X$ for some circuit $C$
(with domain of size $\Lambda$). The reason is that this would not discard
assignments in the image of $C$: It is not necessarily true that $C(\vec
W)\not\equiv X \models X\not\equiv \vec{b}$ for all $\vec b \in \BB^n$ such
that $b\in {\Im}(C)$ (notice that even for such a $\vec b$ there might be
some assignment $\vec a$ for which $C(\vec a)\not\equiv \vec b$).

On the other hand, Jan Kraj\'i\v{c}ek~\cite{Kra07-pers} observed that it is possible to
discard assignments by an axiom of the form $C(\vec{W})\not\equiv X$, where $C$
is a fixed circuit with domain of size at most $\Lambda$ (that is, it has $k<n$
number of input bits, where $2^k\leq \Lambda$), and where the rule of using
this axiom is that we can introduce any instance of $C(\vec{W})\not\equiv X$
where \emph{all the variables in $\vec W$ are substituted by constants $0,1$
and variables from $X$}. This choice of axioms simplifies somewhat the actual
formulation of the promise axioms, as it does not require that $C$ computes an
injective Boolean map. However, a possible drawback of such a formulation is
the following: It is possible that for certain circuits (of the appropriate
number of input and output bits) we shall need to use exponentially many such
axiom instances  to discard \emph{all} (or most of) the assignments
pertaining to the image of the circuits. In contrast to this, our formulation
of the promise axioms above enables a single instance of a promise axiom using
\emph{any} circuit (more correctly, a sequence of circuits of the appropriate
number of input and output bits) to discard \emph{all} the assignments outside
the image of the circuit.

\end{comment}

\section{Big Promise: Upper Bound}\label{secBigPro}

In this section, we show that under the promise $\Lambda=\eps\cd2^n\,$, for any
constant $0<\eps<1$, resolution can efficiently certify the unsatisfiability of
all unsatisfiable 3CNF formulas. The proof method resembles the algorithm
presented by Trevisan~\cite{Tre04}. For a constant $k$, this algorithm receives a $k$CNF
formula $K$ and deterministically approximates the fraction of satisfying
assignments of $K$ within an additive error of $\eps$. The running time of the
algorithm is linear in the size of $K$ and polynomial in
$1/\eps$. 

The idea behind the refutations in this section is based on the following
observation: Given an unsatisfiable 3CNF formula $K$ and
a  constant $c$, either there are $3(c-1)$ variables that hit%
\footnote{A set of variables $S$ that ``hit all the clauses in a CNF
formula $K$'' is a set of variables for which every clause in $K$ contains some
variable from $S$.}
all
the clauses in $K$ or there are at least $c$ clauses in $K$ over $3c$
\emph{distinct} variables denoted by $K'$ (that is, each variable in $K'$
appears only once). In the first case, we can consider all the possible truth
assignments to the $3c$ variables inside resolution: if $K$ is unsatisfiable
then any such truth assignment yields an unsatisfiable 2CNF formula, which can
be efficiently refuted in resolution (cf.~\cite{Coo71}). In the second case, we
can make use of a promise axiom to efficiently refute $K'$ (this set of clauses
has less then $\Lambda$ satisfying assignments, for sufficiently large $c$).
Specifically, in the second case, we construct a sequence of small circuits $C$
for which any satisfying assignment for $K'$ is \emph{provably in resolution}
(with polynomial-size proofs) outside the image of $C$.

The following is the main result of this section:

\begin{theorem}\label{thmBig}
Let $\,0<\eps<1$ be a constant and let $\Lambda=\eps\cd 2^n$ be the given
promise. Then \emph{every} unsatisfiable $3$CNF with $n$ variables has a
polynomial-size (in $n$) resolution refutation under the promise $\Lambda$.
\end{theorem}

This theorem is a consequence of the three lemmas that follow.


\begin{lemma}\label{lemBig1}
Let $K$ be a $3$CNF formula. For every integer $c$ one of the following holds:
(i) there is a set of at most $3(c-1)$ variables that hit all the clauses in $K$; or
(ii) there is a sub-collection of clauses from $K$, denoted $K'$, with at least $c$ clauses and
where each variable appears only once in $K'$.
\end{lemma}

\begin{proof} Assume that $c>2$ (otherwise the lemma is trivial).
Suppose that there is no set of at most $3(c-1)$ variables that hit all the
clauses in $K$ and let $D_1$ be some clause in $K$. Then, there ought to be a
clause $D_2$ from $K$ that contains $3$ variables that are not already in $D_1$
(or otherwise, the $3$ (distinct) variables in $D_1$ hit all the clauses in
$K$, which contradicts the assumption). In a similar manner we can continue to
add new clauses from $K$ until we reach a set of $c$ clauses
$D_1,D_2,\ldots,D_c$, where no variable appears more than once in this set of
clauses.
\end{proof}


If case (i) of the prior lemma holds, then the following lemma suffices to efficiently
refute the 3CNF:

\begin{lemma}\label{lemBig2}
Let $c$ be constant and $K$ be an unsatisfiable $3$CNF formula in the $X$
variables (where $n=|X|$). Assume that there is a set $S\subseteq X$ of at most
$3(c-1)$ variables that hit all the clauses in $K$. Then there is a
polynomial-size (in $n$) resolution refutation of $K$.
\end{lemma}

\begin{proofsketch}
We simply run through all  truth assignments to the variables in $S$ (since
$|S|\leq 3(c-1)$, there are only constant number of such truth assignments).
Under each truth assignment to the $S$ variables, $K$ becomes an unsatisfiable
2CNF. It is known that any unsatisfiable 2CNF has a polynomial-size resolution
refutation (cf.~\cite{Coo71}). Thus, we can refute $K$ with a polynomial-size
resolution refutation.
\end{proofsketch}\medskip

If case (ii) in Lemma~\ref{lemBig1} holds, then
it suffices to show that resolution under a big promise can
efficiently refute any 3CNF formula $T$ with a constant number of clauses (for
a sufficiently large constant), where \emph{each variable in $T$ occurs only
once} (such a $T$ is of course satisfiable, but it has less than an $\eps$
fraction of satisfying assignments for a sufficiently large number of clauses).
This is established in the following lemma.


\begin{lemma}\label{lemBig3}
Fix the constant $c=3\lceil\log_{7/8}(\eps/2)\rceil$.
Let $\Lambda=\eps\cd2^n$, where $0<\eps<1$ is a constant and $n$ is
sufficiently large. Assume that $T$
is a $3$CNF with $c/3$ clauses (and $c$ variables) over the $X$ variables,
where each variable in $T$ occurs only once inside $T$. Then there is a
polynomial-size resolution refutation of $\,T$ under the promise $\Lambda$.
\end{lemma}

\begin{proof}

The proof consists of constructing a sequence of
polynomial-size circuits $C$ (where the parameters of the circuits in $C$ are
taken from Definition~\ref{defAx}; that is, $r=\lceil\log(1/\eps)\rceil$ and
$t=2^r-1$), such that: (i) resolution can efficiently prove the injectivity and
the pairwise disjointness of the images of the circuits in $C$; and (ii) there
is a polynomial-size refutation of $T$ and PRM$_{\Lam, C}$.  In other words,
there is a polynomial-size derivation of the empty clause from the clauses of
both $T$ and PRM$_{\Lam,C}$.

Without loss of generality we assume that the variables in $T$ are
$x_1,\ldots,x_{c}$. The sequence $C$ consists of the circuits $C^{(1)},\ldots,
C^{(t)}$, where each circuit $C^{(i)}$ has $n-r$ input bits and $n$ output
bits. Denote the Boolean circuit that computes the $j$th output bit of
$C^{(i)}$ by $C^{(i)}_j$ and let the input variables of all the circuits in $C$
be $\vec W:=\set{w_1,\ldots, w_{n-r}}$. As shown in equation (\ref{eqImg}),
since the circuits in $C$ are intended to compute injective and pairwise
image-disjoint maps, the image of $C$ would be of size $2^n-2^{n-r}$. We now
define the map that each circuit in $C$ computes.

First, we determine the Boolean functions computed by the output bits in
positions $c+1,\ldots, n$ in all the circuits in $C$. For all $1\leq i\leq t$
and all $c+1 \leq j\leq n$ let $C^{(i)}_j(\vec W)$ compute the $(j-r)$th input
variable $w_{j-r}$.

Second, we need to determine the rest of the output bits for all the circuits
in $C$, that is, we need to determine the Boolean functions computed by
$C^{(i)}_j$, for all $1\leq i\leq t$ and all $1\leq j\leq c$.  Our intention is
that for all $1\le i\le t$, the (single output) circuits
$C^{(i)}_1,\ldots,C^{(i)}_{c}$ should compute (when combined together) a
Boolean map, denoted by $f_i$, from $c-r$ input bits $\,\vec
W_0:=\set{w_1,\ldots,w_{c-r}}$,\, to $c$ output bits. The $j$th output bit of
$f_i$ (which is computed by $C^{(i)}_j$) is denoted by $f_{i,j}$, for $1\le
j\le c$. In other words, $f_i(\vec W_0)=f_{i,1}(\vec W_0)\circ\cdots\circ
f_{i,c}(\vec W_0)$, where $\circ$ denotes concatenation of bits (we shall
describe the functions $f_i$ below). Summing it up for now, we have the
following:

\begin{equation}\label{eqTab}
\begin{array}{l}
 C_1^{(1)} (\vec W_0 ) = f_{1,1} (\vec W_0 ),~ \ldots ,~
		   C_c^{(1)} (\vec W_0 ) = f_{1,c} (\vec W_0 ),\\
		   \qquad \qquad C_{c + 1}^{(1)} (w_{c - r + 1} ) = w_{c - r + 1} , ~\ldots ,~C_n^{(1)} (w_{n - r} ) = w_{n - r}  \\
\hspace{2cm} \vdots \hspace{4cm}  \vdots  \\
 C_1^{(t)} (\vec W_0 ) = f_{t,1} (\vec W_0 ), ~\ldots ,~
		C_c^{(t)} (\vec W_0 ) = f_{t,c} (\vec W_0 ),\\
		\qquad \qquad C_{c + 1}^{(t)} (w_{c - r + 1} ) = w_{c - r + 1} ,
		~\ldots ,~C_n^{(t)} (w_{n - r} ) = w_{n - r},  \\
 \end{array}
\end{equation}

\ind where $C_{j}^{(i)} (w_{k}) = w_{k}$ denotes the fact that $C_{j}^{(i)}$
outputs the (input) variable $w_{k}$ (in which case we assume that the circuit
$C_{j}^{(i)}$ consists of only a single gate: the variable $w_{k}$); and where
$C_{j}^{(i)} (\vec W_0) = f_{i,j}(\vec W_0)$ denotes the fact that
$C_{j}^{(i)}$ computes the function $f_{i,j}$ in the $c-r$ input variables
$\vec W_0$.

We now describe the requirements from the functions $f_{i,j}$. 
Specifically, let $B\subseteq\BB^{c}$ be the set of all \emph{falsifying}
assignments%
\footnote{Note that the assignments here are actually
\emph{partial} truth assignments with respect to $X$, that is, they give truth
values only to the variables $x_1,\ldots,x_{c}$ (these are all the variables in
$T$).}
to $T$ and denote by $\Im(f_i)$ the image of $f_i$,
for all $1\le i\le t$.
We need the $f_i$'s functions to map every input (over $c-r$ input bits)
to a truth assignment (over the $c$ variable $x_1,\ldots,x_{c}$) that
\emph{falsifies} $T$ (that is, a truth assignment from $B$). We also need the
$f_i$'s to be injective and have pairwise disjoint images by the requirements
of the promise axiom (Definition~\ref{defAx}). So that, overall, the Boolean
maps computed by the circuits in $C$ would \emph{discard all the truth
assignments that satisfy $T$}. To prove the existence of such $f_i$'s we need
the following claim:

\begin{claim}\label{claExs} There exists a collection of Boolean functions $f_{i}$,
where $1\leq i\leq t$, for which the following three properties hold.
\begin{enumerate}
  \item $\cup_{i=1}^t${\rm Im}$(f_i)\subseteq B$; \label{item1}
  \item All the $f_{i}$'s are injective and have pairwise disjoint images;
  \item All the $f_{i}$'s depend on a constant number of input variables: $w_1,\ldots,w_{c-r}$
  (and hence, all the $f_{i,j}$'s depend only on these variables).
\end{enumerate}
\end{claim}

\begin{proof}
Each $f_{i}$ should depend on $c-r$ variables and should be injective, and further, each pair of
$f_i$'s should have disjoint images; thus we have:

\begin{equation}\label{eqBig05}
\left|\bigcup\limits_{i=1}^t{\Im}(f_i)\right|=t\cd 2^{c-r}=(2^r-1)\cd 2^{c-r}=2^c  \cdot (1
- 2^{ - r} )\,.
\end{equation}

Hence, to prove the existence of the collection of $f_{i}$'s with the required three properties
it suffices to show that $|\cup_{i=1}^t{\Im}(f_i)| \le |B|$, and by (\ref{eqBig05}) it
suffices to show:

\begin{equation}\label{eqBig10}
  2^c  \cdot (1 - 2^{ - r} ) \le |B|\,.
\end{equation}

Observe that the fraction of distinct assignments  that satisfy $T$ is equal to
the probability (over all truth assignments to $T$) that a uniformly chosen
random truth assignment satisfies all the $c/3$ clauses in $T$, which is equal
to

\begin{equation}\label{eqFrac}
\left( {\frac{7}{8}} \right)^{c/3}  =
\left(\frac{7}{8}\right)^{\lceil\log_{7/8}(\eps/2)\rceil},
\end{equation}
and so
$$|B| = 2^c  \cdot \left( {1 - \left( {\frac{7}{8}} \right)^{\left\lceil {\log _{7/8}
(\eps /2)} \right\rceil } } \right).$$

Therefore, for (\ref{eqBig10}) to hold it remains to show

$${\rm{ }}2^{ - r}  \ge \left( {\frac{7}{8}} \right)^{\left\lceil {\log _{7/8} (\eps /2)}
\right\rceil }, $$
which holds because
\[
\begin{array}{c}
2^{ - r}  = 2^{ - \left\lceil {\log (1/\eps )} \right\rceil }  \ge 2^{ - \log (1/\eps ) -
1}  = 2^{ - \log (1/\eps )} /2 \\= \eps /2 =
\left( {\frac{7}{8}} \right)^{\log
_{7/8} (\eps /2)}  \ge \left( {\frac{7}{8}} \right)^{\left\lceil {\log _{7/8}
(\eps /2)} \right\rceil }.
\end{array}
\]
\end{proof}

Having established the existence of functions $f_{i}$ for which the three
conditions in Claim~\ref{claExs} hold, we define each circuit $C^{(i)}_j$, for
$1\leq i\leq t$ and $1\leq j\leq c$, to compute the function $f_{i,j}$. Since
the domain of each $f_{i,j}$ is constant (that is, $2^{c-r}$) \emph{each
$C^{(i)}_j$ can be of constant size}.

Note that the circuits in $C=(C^{(1)},\ldots,C^{(t)})$ indeed compute $t$
injective Boolean maps that have pairwise disjoint images. Disjointness of
images stems from the fact that the $f_i$'s functions all have disjoint images,
and injectivity stems from the fact that the $f_i$'s are all injective, and
that for each $1\le i\le t$, the Boolean map computed by
$C^{(i)}_{c+1}\circ\ldots\circ C^{(i)}_{n}$ (again, $\circ$ denotes
concatenation of bits) is exactly the identity map
${\rm{id}}\!:\BB^{n-c}\to\BB^{n-c}$.

To complete the proof of Lemma~\ref{lemBig3}, we need to show that
\emph{resolution can efficiently prove }that indeed the circuits in $C$ all
compute injective Boolean maps and have pairwise disjoint images (as well as to
efficiently refute $T$ when assuming that $X$ can take only assignments from
the image of $C$). This is done in the following claim:

\begin{claim}\label{claProvability} Let $C$ be the sequence of circuits as devised above, and
let $\,{\rm{PRM}}_{C,\Lambda}$ be the corresponding (big) promise axiom. Then
there is a polynomial-size resolution refutation of $\,T$ and
${\rm{PRM}}_{C,\Lambda}$.
\end{claim}

\begin{proof}
The proof follows by considering the encoding of the (big) promise axiom
$\,{\rm{PRM}}_{C,\Lambda}$ via the encoding scheme in the Appendix (Section
\ref{secEnc}) and showing how resolution can prove the empty clause from $T$
and this encoding. Here we shall use a less formal description; more details
can be found in the appendix.

First we need resolution to prove the premise of the main implication in
${\rm{PRM}}_{C,\Lambda}$. This breaks into two parts corresponding to $
\wedge_{i=1}^{t} \left(C^{(i)}(\vec{W_1})\equiv C^{(i)}(\vec{W_2}) \to
\vec{W_1} \equiv \vec{W_2}\right)$\, and $\,\wedge_{1\le i< j\le t}
C^{(i)}(\vec{W_1})\not\equiv C^{(j)}(\vec{W_2})$.

For the first part, we need to refute the statement expressing that $C$ contains
some circuit $C^{(i)}$ that computes a non-injective Boolean map. This can be
efficiently refuted in resolution: Assume (inside resolution) that $C^{(i)}
(\vec W_1 ) \equiv C^{(i)} (\vec W_2 )\,$, for some $1\le i\le t$, then by
(\ref{eqTab}) we can efficiently prove (inside resolution) that for all $c -r+1
\le j\leq n-r\,$ it happens that $w^{(1)}_j \equiv w^{(2)}_j$ (where
$w^{(1)}_j$ is the $j$th variable in $\vec W_1$, and $w^{(2)}_j$ is the $j$th
variable in $\vec W_2$) (see details in the appendix, and in particular Section
\ref{secInsideRes}). Thus, it remains to refute the statement that for some
$1\le j\le c-r$ it happens that $w^{(1)}_j \not\equiv w^{(2)}_j$. This is
indeed a contradiction by definition of the circuits in $C$ (as they compute
injective maps). Since all the output bits $w^{(1)}_j, w^{(2)}_j$ for $1\le
j\le c-r$, are computed by constant size circuits $C^{(i)}_j$ for $1\le j\le
c-r$ and $1\le i\le t$ (with constant number of input variables), such a
contradiction can be refuted in constant size resolution refutation (again, see
more details in the appendix).

The disjointness of the images of the (maps computed by the) circuits in $C$
is also efficiently provable inside resolution in a similar manner, and we
shall not describe it here.

Therefore, we arrive (inside resolution) at the consequence of the main
implication in the promise axiom: $\vee_{i = 1}^t {C^{(i)} (\vec W_1 )
\equiv X} $. It remains to refute $T$ and $\vee_{i = 1}^t {C^{(i)} (\vec W_1
) \equiv X} $.

Again, $T$ has a constant number of variables (that is, $c$). 
Consider only the circuits that output to the variables $x_1,\ldots,x_c$ in
$\vee_{i = 1}^t {C^{(i)} (\vec W_1 ) \equiv X}$: these are the circuits
$C^{(i)}_1,\ldots,C^{(i)}_c$ for all $1\le i\le t$. We shall denote the set of
these circuits by $C'$. The (functions computed by) the circuits in $C'$ depend
on a constant number of variable $\vec W_0$ and they have constant size. Denote
by $Z$ the subformula of PRM$_{C,\Lambda}$ that contains the (encoding of) the
circuits in $C'$ including the encoding of the statement that for some $1\le
i\le t$ the variables $x_1,\ldots,x_c$ are equal to the output of the circuits
$C^{(i)}_1,\ldots,C^{(i)}_c$. By the definition of the circuits in $C$ (see
condition (\ref{item1}) in Claim~\ref{claExs}) $Z$ discards all the satisfying
assignments of $T$ (over the $c$ variables in $T$)\footnote{We
have abused notation here, as we defined \emph{assignment discarding} of only
\emph{complete assignments to $X$}
while here we say that $Z$ discards a
partial assignment to the variables $x_1,\ldots,x_c$ only; but the definition
of such partial assignment discarding is similar (consider the variables
$x_1,\ldots,x_c$ to be the only \emph{original variables}).}. Thus, $T$ and $Z$
constitute together a contradiction of constant size (as there are no
satisfying assignments for both $T$ and $Z$). Therefore, there is a constant
size resolution refutation of $T$ and $Z$.
\end{proof}

This concludes the proof of Lemma~\ref{lemBig3}.
\end{proof}

\section{Smaller Promise: Lower Bound}\label{secSmlPro}
In this section, we prove an exponential lower bound on the size of resolution
refutations under the promise $2^{\delta n}$, for any constant $0 \le \delta
\le 1$. The lower bound apply to random 3CNF formulas with $\SmallO(n^{3/2})$
number of clauses (where $n$ is the number of variables in the 3CNF). This
lower bound matches the known lower bound on resolution refutation-size for
random 3CNF formula (without any promise). Basically, the proof strategy of our
lower bound is similar to that of Ben-Sasson and Wigderson~\cite{BSW99}, except that we need to take
care that every step in the proof works with the augmented (smaller) promise
axiom.

The lower bound is somewhat stronger than described above in two respects.
First, we show that restricting the set of all truth assignments $2^n$ to
\emph{any} smaller set (that is, not just those sets defined by small circuits)
that consists of $2^n-2^{\delta n}$ assignments (for any constant $0 \le
\delta\le 1$), does not give resolution any advantage in the average-case. One
can think of such a restriction as modifying the semantic implication relation
$\models$ to take into account only assignments from some prescribed set of
assignments $S$, such that $|S|=2^n-2^{\delta n}$
(in other words, for two formulas $A,B$, we have that $A\models B$
under the restriction to $S$ iff any truth assignment \emph{from $S$} that
satisfies $A$ also satisfies $B$).
 Formally, this means that the lower bound does not use the fact that
the restricted domain of size $2^n-2^{\delta n}$ is defined by a sequence
$C$ of polynomial-size
circuits (nor the fact that the circuits in $C$ ought to have polynomial-size
resolution proofs of their injectivity and pairwise disjointness).

Second, we could allow for a promise that is bigger than $2^{\delta n}$, and in
particular for a promise of $\,2^{n(1-1/n^{1-\xi})}=2^n/2^{n^{\xi}}$, for some
constant $0<\xi<1$ (see Remark~\ref{remLB} below). The actual proof of the
lower bound uses the smaller promise of $2^{\delta n}$, but the proof for a
$2^n/2^{n^{\xi}}$ promise is the same. (Although we have not defined precisely
how the promise axioms are formulated in the case of a promise equal to
$2^n/2^{n^{\xi}}$, it is possible to formulate such promise axioms along the
same lines described in Definition~\ref{defAx2}.)

The following defines the usual \emph{average-case} setting of 3CNF formulas
(there are other definitions, that are essentially similar):

\begin{definition}[{Random $3$CNF formulas}]\label{defRan3CNF}
For a $3${\rm{CNF}} formula $K$ with $n$ variables $X$ and $\beta\cd n$
clauses, we say that $\beta$ is the \emph{density} of $K$. A \emph{random
$3${\rm{CNF}} formula} on $n$ variables and density $\beta$ is defined by
picking $\beta\cd n$ clauses from the set of all $\,2^3\cd {n \choose
3}$ clauses, independently and indistinguishably distributed, with repetitions.
\end{definition}

We say that an event (usually a property of a 3CNF in $n$ variables and density
$\beta$) happens \emph{with high probability} if it happens with $1-o(n)$
probability in the specified probability space (usually random $3$CNF formulas
as defined in Definition~\ref{defRan3CNF}).

Our goal is to prove a lower bound on the average-case refutation-size of 3CNF
formulas taken from the \emph{set of 3CNF formulas under a promise} as defined
in Definition~\ref{defCNFpro} (note that the probability space defined in
Definition~\ref{defRan3CNF} is defined over a different set of 3CNF, that is, the
set of \emph{all} 3CNF formulas). For this purpose, we define a probability
space over the set of 3CNF formulas under a promise: The distribution of
\emph{random $3${\rm{CNF}} formulas under a promise $\Lambda$} on $n$ variables
and density $\beta$ is the distribution of random $3${\rm{CNF}} formulas in
Definition~\ref{defRan3CNF} \emph{conditioned} on the event that the
$3${\rm{CNF}} is either unsatisfiable or has more than $\Lambda(n)$ satisfying
assignments.

We now argue that to satisfy our goal to prove a lower bound on the
average-case proof complexity of 3CNF formulas under a promise, it is
sufficient to prove the lower bound result considering the distribution of
random 3CNF formulas as defined in Definition~\ref{defRan3CNF}.

It is well known that almost all 3CNF formulas with a density $\beta$ above a
certain constant threshold (say, $5$) are unsatisfiable. This means that any
property of a 3CNF (with density above the threshold) that happens with high
probability in the distribution in Definition~\ref{defRan3CNF} also happens
with high probability in the distribution of random $3${\rm{CNF}} formulas
under a promise $\Lambda(n)$ (as defined above); this is because there are only
a fraction $\SmallO(1)$ of 3CNF formulas (with a given fixed number of
variables $n$ and a given fixed density $\beta$ above the threshold) that are
satisfiable (and moreover have at least one satisfying assignment but less than
$\Lambda(n)$ satisfying assignments). Thus, if we prove that with high
probability a random 3CNF formula has no small promise resolution refutation
then it implies also that with high probability \emph{a random 3CNF formula
under a promise} has no small promise resolution refutation. \emph{Therefore,
we shall consider from now on only the distribution of 3CNF formulas as defined
in Definition~\ref{defRan3CNF}, and forget about the other distribution}.

\subsection{The Lower Bound}\label{secRanCNF}

Throughout this section we fix $0<\delta<1$ and $\Lambda= 2^{\delta n}$. We
also fix an arbitrary instance of a promise axiom PRM$_{C,\Lambda}$ (from
Definition~\ref{defAx2}; where $C$ is a sequence of the appropriate number of
circuits, and each circuit in $C$ have the appropriate number of input and
output bits). For $K$ a CNF formula, we denote by ${\rm{Vars}}(K)$ the set of
variables that occur in $K$.

The following is the main theorem of this section. The lower bound matches that
appearing in~\cite{BSW99} for resolution. 

\begin{theorem}\label{thmSma} Let $0<\delta<1$ and $0<\epsilon< 1/2$. With high
probability a random $3$CNF formula with $\beta=n^{1/2-\epsilon}$ requires a
size $\,\exp (\mathrm\Omega (\beta ^{ - 4/(1 - \epsilon )} \cdot n))$ resolution
refutation under the promise $\Lambda=2^{\delta n}$.
\end{theorem}

\begin{remark}\label{remLB}
As mentioned above, we could allow in Theorem~\ref{thmSma} for a promise that
is bigger than $2^{\delta n}$, and precisely for a promise of
$\,2^{n(1-\frac{1}{n^{1-\xi}})}=2^n/2^{n^{\xi}}$, for any  constant $\xi$ such
that $\frac{\epsilon}{(1-\epsilon)}<\xi<1$ (for instance, this allows for a
promise of $2^n/2^{n^{1/3}}$).
\end{remark}

The proof strategy of Theorem~\ref{thmSma} is to show that with high
probability for a random 3CNF formula $K$ with density
$\beta=n^{1/2-\epsilon}$, a resolution refutation under the promise $2^{\delta
n}$ of $K$ must contain some clause $D$ of large width. Then we can apply the
size-width tradeoff from Theorem~\ref{thmBSW} to reach the appropriate size
lower bound.

However, we need to be a bit careful here, as in order to illustrate an
exponential lower bound via the size-width tradeoff of Theorem~\ref{thmBSW}, we
need to guarantee that all the initial clauses (that is, all the axiom clauses)
are of \emph{constant width}. The 3CNF formula $K$ is certainly of constant
width, but the clauses pertaining to the promise axiom PRM$_{C,\Lambda}$ might
not be (see the appendix for a detailed specification of these clauses). We can
solve this problem easily: First, we add yet more extension variables to encode
the clauses of the promise axiom with new constant width clauses. Second, we
note that the original clauses of the promise axiom can be derived by a
linear-size resolution proof from the new constant width version of the promise
axiom (therefore, if there is a polynomial-size resolution refutation of $K$
using the original promise axiom, then there is also a polynomial-size
resolution refutation of $K$ using the new constant width version of the
promise axiom). Finally, we prove the exponential lower bound on resolution
augmented with the constant width version of the promise axiom (instead of the
original clauses pertaining to the promise axiom).

Let us  explain now how to get the new constant-width promise axiom from the
clauses pertaining to the original promise axiom from Definition~\ref{defAx2}
(as depicted in the appendix). Let $E=\ell_1\vee\ldots\vee \ell_m$ be a clause
in the promise axiom that has more than constant width (that is, $\ell_i$'s are
literals and $m=\omega(1)$). Then, we replace the clause $E$ with the following
collection of clauses:

\begin{equation}\label{eqClauses}
\mbox{$\,\ell_1 \vee e_1$, $\neg e_1\vee \ell_2\vee e_2$, $\,\neg e_2\vee
\ell_3\vee e_3,\,\ldots\,,\neg e_{m-1}\vee \ell_m$},
\end{equation}
where the $e_i$'s are new extension variables. By resolving on the $e_i$
variables, one after the other, it is possible to derive with a linear-size
resolution proof the original clause $E$ from the clauses in (\ref{eqClauses})
(consider the first two clauses (from left) in (\ref{eqClauses}), and resolve
over the variable $e_1$, then the resolvent of this step is resolved over the
variable $e_2$ with the third clause in (\ref{eqClauses}), and so forth). (Note
that every truth assignment that satisfies (\ref{eqClauses}) also satisfies
$E$, and so any clause that is semantically implied by $E$ (see the
preliminaries, Section~\ref{secPre}) is also semantically implied by
(\ref{eqClauses}). This means that the new constant width version of the
promise axiom discards the same truth assignments to the variables $X$ as the
original version of the promise axiom.)

Thus, \emph{from now on in this section we assume that the promise axiom
consists of clauses of a constant width.}

The rest of this section is devoted to the proof of Theorem~\ref{thmSma}.

For a clause $D$ define:
$$\eta(D):=\min\set{|K'|\; \Big| \;K'\subseteq K\,\,
 {\rm and\,\, (PRM}_{C,\Lambda}\cup K')\models D}.$$
\begin{remark}
We use the symbol $\eta$ to distinguish it from a similar measure $\mu$ used in
\cite{BSW99}: Here we require the minimal set of clauses from $K$ that
\emph{combined with the axiom {\rm{PRM}}$_{C,\Lambda}$} semantically imply $D$.
\end{remark}
We show that with high probability for a random 3CNF formula with density
$\beta=n^{1/2-\epsilon}$, for $0<\epsilon<1/2$, the following is true:
\begin{enumerate}
  \item Let $k=2n \cd(80\beta)^{-2/(1-\epsilon)}$. Then $\eta(\Box) \ge k$.\label{it1}
  \item Any refutation of $K$ must contain a clause $D$ such that
   $k/2 \leq \eta(D)\leq k$.\label{it2}
  \item Any clause $D$ from~\ref{it2} must have large width, and specifically $|D|\ge
\epsilon n(80\beta)^{-2/(1-\epsilon)}$ (which, by Theorem~\ref{thmBSW},
concludes the proof).\label{it3}
\end{enumerate}

The  following two definitions are similar to those in~\cite{BSW99} (we refer
directly to 3CNF formulas instead of $3$-uniform hypergraphs):

\begin{definition}[{CNF Expansion}]\label{defExp} For a 3CNF formula $K$ with density
$\beta=n^{1/2-\epsilon}$, for $0< \epsilon< 1/2$, the \emph{expansion} of $K$
is
\[
e(K):=\min\left\{2|{\rm{Vars}}(K')|-3|K'|\;\Bigg|\;
\begin{array}{l}
K'\subset K \,\mbox{
{\rm{and}} }\\ n\cd(80\beta)^{-2/(1-\epsilon)}\le |K'| \\\le
2n\cd(80\beta)^{-2/(1-\epsilon)}
\end{array}\right\}.
\]
\end{definition}

\begin{definition}[{Partial matchability}] A 3CNF formula $K$ with density
$\beta=n^{1/2-\epsilon}$, for $0< \epsilon< 1/2$, is called \emph{partially
matchable} if for all $K'\subset K$ such that $|K'|\le 2n
\cd(80\beta)^{-2/(1-\epsilon)}$ we have $|{\rm{Vars}}(K')|\ge |K'|$.
\end{definition}

The next lemma gives two properties of random $3$CNF formulas that occur with
high probability (see the appendix of~\cite{BSW99} for a proof). We then use
this lemma to show that with high probability~\ref{it1},\ref{it2},\ref{it3}
above hold.
\begin{lemma}[{\cite{BKPS02}}]\label{lemBeame}
Let $0< \epsilon<  1/2$ and let $K$ be a random 3CNF with $n$ variables and
density $\beta=n^{1/2-\epsilon}$, then with high probability:
\renewcommand{\theenumi}{\arabic{enumi}} 
\begin{enumerate}
  \item  $e(K)\ge \epsilon n(80\beta)^{-2/(1-\epsilon)}$; and\label{itBeame1}
  \item $K$ is partially matchable.\label{itBeame2}
\end{enumerate}
\end{lemma}

\emph{We now prove}~\ref{it1}. In light of part (\ref{itBeame2}) in Lemma
\ref{lemBeame}, in order to prove that with high probability~\ref{it1} holds it
is sufficient to prove the following:
\begin{lemma}\label{lemMatch}
Let $K$ be a $3$CNF formula in the $X$ variables with density
$\beta=n^{1/2-\epsilon}$, for $0< \epsilon< 1/2$. If $K$ is partially matchable
then $\eta(\Box)\ge2n \cd(80\beta)^{-2/(1-\epsilon)}$.
\end{lemma}

\begin{proof}
By partial matchability of $K$, for all $K'\subset K$ such that $|K'|\le 2n
\cd(80\beta)^{-2/(1-\epsilon)}$ it happens that $|{\rm{Vars}}(K')|\ge |K'|$.
Thus, by Hall's Theorem we can choose a distinct variable (representative) from
each clause in $K'$ and set it to satisfy its clause. Clearly,
$|{\rm{Vars}}(K')|\le 3|K'|\le 6n\cd(80\beta)^{-2/(1-\epsilon)}$, and so there
is a (partial) truth assignment $\rho$ to at most
$6n\cd(80\beta)^{-2/(1-\epsilon)}$ variables in $X$ that satisfies $K'$. Since
$\beta=n^{1/2-\epsilon}$,

\begin{equation}\label{eqBeame}
6n\cd(80\beta)^{-2/(1-\epsilon)}=6\cd 80^{-2/(1-\epsilon)}\cd
n^{\epsilon/(1-\epsilon)},
\end{equation}
which, by $0<\epsilon< 1/2$, is equal to $O(n^{\lambda})$ for some
$0<\lambda<1$. Thus for sufficiently large $n$ there are more than $\delta n$
variables from $X$ not set by $\rho$, which means that there are more than
$2^{\delta n}$ different ways to extend $\rho$ into truth assignments (to all
the variables in $X$) that satisfy $K'$.\footnote{Actually,
for sufficiently large $n$ there are more than
$\Omega(n-n^{\epsilon/(1-\epsilon)})$ such variables, from which we can assume
the bigger promise $\Lambda=2^n/2^{n^\xi}$, for any
$\frac{\epsilon}{(1-\epsilon)}<\xi<1$, as noted in Remark~\ref{remLB}.} Since
the promise axiom PRM$_{C,\Lambda}$ can discard up to $2^{\delta n}$ truth
assignments to the $X$ variables, we get that PRM$_{C,\Lambda}\cup K'$ is
\emph{satisfiable} (any assignment to $X$ that is not discarded by
PRM$_{C,\Lambda}$ can be extended to the extension variables in a way that
satisfies PRM$_{C,\Lambda}$).

We have thus showed that every collection $K'$ containing at most $2n
\cd(80\beta)^{-2/(1-\epsilon)}$ clauses from $K$ and augmented with the promise
axiom PRM$_{C,\Lambda}$ is satisfiable. This implies in particular that
$\eta(\Box)\ge 2n \cd(80\beta)^{-2/(1-\epsilon)}$.
\end{proof}

\emph{We now prove}~\ref{it2}. Note that the resolution rule is
\emph{sub-additive} with respect to $\eta$ in the sense that for all three
clauses $E,F,D$ such that
 $D$ is a resolvent of $E$ and $F$, it holds that
$$\eta(E)+\eta(F)\ge\eta(D).$$
We also clearly have that for every axiom clause $E$ (either from $K$ or from
the promise axiom):
$$\eta(E)=1.$$
Let $$k=2n \cd(80\beta)^{-2/(1-\epsilon)}.$$ By Lemma~\ref{lemMatch}, with high
probability for a 3CNF formula $K$ with density $\beta=n^{1/2-\epsilon}$ (for
$0< \epsilon< 1/2$) it happens that $\eta(\Box)\ge k$. By sub-additivity of the
resolution rule with respect to $\eta$, in any resolution refutation of $K$
under the promise $\Lambda$, there ought to be some clause $D$ such that
$$k/2\leq \eta(D)\leq k.$$

\emph{We now prove}~\ref{it3}. Let $D$ be a clause such that $k/2\leq
\eta(D)\leq k$ from~\ref{it2} and let $K'$ be the (minimal) set of clauses from
$K$ for which PRM$_{C,\Lambda}\cup K'\models D$ and $k/2 \leq |K'|\leq k$. We
shall prove that (with high probability for a random 3CNF) $|D|\geq \epsilon
n(80\beta)^{-2/(1-\epsilon)}$. In light of Lemma~\ref{lemBeame} part
(\ref{itBeame1}), in order to prove this, it is sufficient to prove the
next two lemmas.

Define $\partial K'$, called the \emph{boundary of $K'$}, to be the set of
variables in $K'$ that occur only \emph{once} in $K'$ (in other words, each
variable in $\partial K'$ appears only in one clause in $K'$).

\begin{lemma}
$|\partial K'|\ge e(K)$.
\end{lemma}

\begin{proof}
Every variable not in $\partial K'$ must be covered by at least two distinct
clauses in $K'$, and so $|{\rm{Vars}}(K')|\le |\partial K'| +
\frac{1}{2}\cd(3|K'| - |\partial K'|)$. Thus, we have $|\partial K'| \ge
2|{\rm{Vars}}(K')|-3|K'| \ge e(K)$ (where the last inequality is by Definition
\ref{defExp} and since $k/2 \leq |K'|\leq k$).
\end{proof}

\begin{lemma}\label{lemClause}
$|D|\geq |\partial K'|$.
\end{lemma}

\begin{proof}
Let $x_i\in \partial K'$, for some $1\le i\le n$, and denote by $K_i$ the (unique)
clause from $K'$ that contains $x_i$. Assume by a way of contradiction that
$x_i$ does not occur in $D$.

By minimality of $K'$ with respect to $\eta$ and $D$ there exists an assignment
$\alpha$ (here we treat $\alpha$ as a \emph{total }truth assignment, that is, a
truth assignment to both the $X$ variables and the extension variables in the
promise axiom) such that
\begin{equation}\label{eqBoundary}
(K'\setminus K_i)(\alpha)=1\,\,\mbox{ and }\,\, D(\alpha)=0
\end{equation}
(as otherwise $\,(K'\setminus K_i) \models D$ which clearly implies
\,PRM$_{C,\Lambda}\cup (K'\setminus K_i) \models D$, which then contradicts the
minimality of $K'$ with respect to $\eta$ and $D$).

By assumption, $x_i$ occurs neither in $K'\setminus K_i$ nor in $D$. Hence, we
can flip the value of $\alpha$ on  $x_i$ so that $K_i(\alpha)=1$ while still
keeping (\ref{eqBoundary}) true. We thus have:

\begin{equation}\label{eqFlip}
K'(\alpha)=1\,\,\mbox{ and }\,\, D(\alpha)=0
\end{equation}

Since $|K'|\le k$, we have that $|{\rm{Vars}}(K')|\le
3k=6n\cd(80\beta)^{-2/(1-\epsilon)}$ (recall that $|K'|$ is the number of
\emph{clauses} in $K'$). If $|D|\geq |\partial K'|$ we are done. Otherwise,
\[
|{\rm{Vars}}(K')|+|D| < |{\rm{Vars}}(K')|+|\partial K'| \le
2|{\rm{Vars}}(K')|\le 12n\cd(80\beta)^{-2/(1-\epsilon)}.
\]
Thus, similar to equation (\ref{eqBeame}), for sufficiently large $n$, the
total number of distinct variables in $K'$ and $D$ is at most
$|{\rm{Vars}}(K')|+|D|=O(n^{\lambda})$, for some $0<\lambda<1$. This means that
for sufficiently large $n$ there are more than $\delta n$ variables from $X$
for which flipping the value of $\alpha$ on them still validates
(\ref{eqFlip}).\footnote{Again, similar to what was noted in
the proof of Lemma~\ref{lemMatch}, for sufficiently large $n$ there are
actually more than $\Omega(n-n^{\epsilon/(1-\epsilon)})$ such variables.}
Hence, there are more than $2^{\delta n}$ distinct assignments to the $X$
variables for which (\ref{eqFlip}) holds.

The promise axiom PRM$_{C,\Lambda}$ discards at most $2^{\delta n}$ assignments
to the $X$ variables. This means that there are at most $2^{\delta n}$
assignments $\rho$ to the $X$ variables that falsify PRM$_{C,\Lambda}$ (that is,
that every extension of $\rho$ to all the extension variables falsifies
PRM$_{C,\Lambda}$), while \emph{all} other assignments $\rho$ to the $X$
variables have an extension (to all the the extension variables) that
\emph{satisfies} PRM$_{C,\Lambda}$. Thus, by the previous paragraph there ought
to be at least one assignment $\rho$ to the $X$ variables that has an extension
$\rho'$ to the extension variables, such that
\begin{equation}\label{eqBoundr2}
{\rm{PRM}}_{C,\Lambda}(\rho')=1\,, K'(\rho')=1\,\,\mbox{ and }\,\, D(\rho')=0,
\end{equation}
which contradicts the assumption that \;PRM$_{C,\Lambda}\cup K'\models D$.
\end{proof}

\section{Conclusion}
This paper establishes a new framework of propositional proof systems that are
able to separate the unsatisfiable CNF formulas from the set of CNF formulas
having many satisfying assignments. We were able to analyze the complexity of
basic cases pertaining to such proof systems, such as the case of a big promise
(a constant fraction of all truth assignments) and the average-case proof
complexity of refutations under a smaller promise (that is, a promise of
$2^{\delta n}$, for any constant $0<\delta<1$).

One question we have not addressed is what can be gained (if at all) when we
augment a stronger proof system than resolution, like bounded-depth Frege proof
system or Frege proof system, with the promise axioms (for a small promise like
$2^{\delta n}$, as for a big promise already
resolution can efficiently refute all unsatisfiable 3CNF formulas).

Another question that arises is whether the fact that we require the Boolean
circuits in the promise axioms to be \emph{provably injective} and to
\emph{provably posses disjoint images} (that is, provably inside resolution)
constitutes a real restriction. (Note that the lower bound for resolution under
the promise $2^{\delta n}$ in Section~\ref{secSmlPro} did not use at all these
requirements.) In other words, we ask whether there is a sequence of circuits
$C^{(1)},\ldots,C^{(t)}$ for which adding the axiom $\vee_{i = 1}^t {C^{(i)}
(\vec W ) \equiv X}$ (where the parameter $t$ and the number of variables in
$m$ are taken from the smaller promise axiom~\ref{defAx2}) to resolution (or a
stronger proof system) gives a super-polynomial speed-up for some contradictory
family of formulas over standard resolution (or the stronger proof system); but
that we cannot prove efficiently in resolution (or the stronger proof system)
that $C^{(1)},\ldots,C^{(t)}$ are injective or that they have pairwise disjoint
images?

A different and more general task is to come up with other natural models of
propositional proof systems that capture a ``relaxed" notion of soundness. For
instance, Pitassi~\cite{PitTalk06} suggested considering ``approximate proofs'' in the
framework of algebraic proof systems.

Finally, we have not dealt directly in this paper with the promise $\Lambda=
2^n/poly(n)$, though it is most likely that a similar upper bound (with a
similar proof) to that shown in Section~\ref{secBigPro} also holds for this
promise (when the promise axiom is modified accordingly). In this respect it is
worth mentioning that Kraj\'i\v{c}ek~\cite{Kra06} observed that the work
of Razborov and Rudich~\cite{RR94}
implies the existence (under a cryptographic conjecture) of a Boolean function
$g$ with $n^\delta$ input bits (denoted by $y_1,\ldots,y_{n^\delta}$) and $n$
output bits (denoted by $g_1(y_1,\ldots,y_{n^\delta}),\ldots,
g_n(y_1,\ldots,y_{n^\delta})$), for any constant $0<\delta<1$, that has the
following property: Given any CNF formula $K$ in $n$ variables $x_1,\ldots,
x_n$, substituting $g_1(y_1,\ldots,y_{n^\delta}),\ldots,
g_n(y_1,\ldots,y_{n^\delta})$ for the original $x_i$ variables in $K$ yields a
new CNF formula that is unsatisfiable only if $K$ has at most
$2^n/n^{\omega(1)}$ satisfying assignments. This means that \emph{under the
promise $2^n/poly(n)$ the substitution $g$ is sound}: Any unsatisfiable CNF
formula (clearly) stays unsatisfiable after the substitution, while any CNF
with more than $2^n/poly(n)$ satisfying assignments stays satisfiable after the
substitution.

\appendix

\section{Encodings}

\subsection{Encoding of Boolean Circuits and Promise Axioms}\label{secEnc}
In this section we describe in detail how the promise axiom (see Definition
\ref{defAx}) is encoded as a CNF formula. As already mentioned in the
Preliminaries (Section~\ref{secPre}), by Cook's Theorem there is always an
efficient way to encode a Boolean circuit as a CNF formula using extension
variables (that is, a CNF with size $O(s\cd\log(s))$ can encode a circuit of
size $s$). However, we shall need to be more specific regarding the actual way
the encoding is done since we require that resolution should be able to
efficiently prove some basic facts about the encoded circuits.

\subsubsection{Boolean circuit encoding} The following definition is similar
to the \emph{circuit encoding} defined in Alekhnovich \emph{et al.} in
\cite{ABSRW00} (note that it deals
with a \emph{single output bit} Boolean circuit):

\begin{definition}[{Encoding of Boolean circuits}]\label{defEnc}
Let $C(\vec{W})$ be a Boolean circuit (with $\vee,\wedge$ as fan-in two gates
and $\neg$ a fan-in one gate) and $m$ input variables $\vec{W}:=w_1,\ldots,w_m$
and a \emph{single output bit}. For every gate $v$ of the circuit $C$ we
introduce a special extension variable $y_v$. For input gates $w_j$ ($1\leq
j\leq m$) we identify $y_{w_j}$ with $w_j$. We denote by $y^1$ the literal $y$
and by $y^0$ the literal $\neg y$. The CNF formula $\|C(\vec{W})\|$ consists of
the following clauses:

(i) $y_{v_1}^{\bar{\epsilon}_1}\vee y_{v_2}^{\bar{\epsilon}_2}\vee
y_v^{\pi_\circ(\epsilon_1,\epsilon_2)}\,$, where $v$ is a
$\,\circ\in\set{\vee,\wedge}$ gate in $C$ and $v_1,v_2$ are the two input gates
of $v$ in $C$ and $\la \epsilon_1,\epsilon_2\ra$ is any vector in $\BB^2$ and
$\pi_\circ$ is the truth table function of $\circ$ (and $\bar 0=1$,\,
$\bar{1}=0$);

(ii) $y_{v_1}^{\bar{\epsilon}_1}\vee y_v^{\pi_\neg(\epsilon_1)}\,$, where $v$
is a $\neg$ gate in $C$ and $v_1$ is the single input gates of $v$ in $C$, and
$\epsilon_1\in\BB$ and $\pi_{\neg}$ is the truth table function of
$\,\neg$.

We write $\|C(\vec{W})\|(y)$ to indicate explicitly that the output gate $v$ of
$C$ is encoded by the extension variable $\,y$.
\end{definition}

\subsubsection{Encoding of the promise axioms} We now give a rather detailed
description of how the promise axioms are encoded as CNF formulas. We shall
consider only the `big' promise axiom (Definition~\ref{defAx}), but the other
variant (Definition~\ref{defAx2}) is similar. We encode the promise axioms in a
bottom-up manner, encoding the sub-formulas separately, and then combining all
of them together.

%
%
%

We assume that a Boolean circuit $C(\vec{W})$ with $n$ output bits is encoded
as $n$ distinct circuits and we write $\|C(\vec{W})\|(\vec{Y})$ to indicate
explicitly that the output gates $v_1,\ldots,v_n$ of $C$ are encoded by the
extension variables $\,y_1,\ldots,y_n$ (where $\vec{Y}:=\set{y_1,\ldots,y_n}$).
This means that $\|C(\vec{W})\|(\vec{Y})$ is the CNF formula $\wedge_{i=1}^n
\|C_i(\vec{W})\|(y_i)$, where $C_i(\vec{W})$ is the circuit computing the $i$th
output bit of $C(\vec{W})$ and $y_i$ is the variable that encodes (see
Definition~\ref{defEnc}) the (single) output bit of $C_i(\vec{W})$. We also
require that if the (function computed by the) circuit $C(\vec{W})$ does not
depend,%
\footnote{We say that a Boolean function $f$ does
not depend on an input bit $w_i$ if for all input assignments $\alpha$ to $f$,
flipping the truth value of $w_i$ in $\alpha$ does not change the value of
$f$.}
 on some input bit $w_i$, then $w_i$ does not occur in the
encoding of $C(\vec{W})$.

%

Let $1\leq k\leq t$ (where the parameter $t$ is taken from Definition
\ref{defAx}). We first encode as a CNF formula the negation of following
sub-formula from the promise axiom:
$$C^{(k)} (\vec W_1 ) \equiv C^{(k)} (\vec W_2 )\, \to \vec W_1 \equiv \vec
W_2\,.$$
 We denote this CNF encoding by $\neg$INJ$_k$ (where INJ stands for
\emph{injective}).

\begin{definition}[{$\neg$INJ$_k$}]\label{defINJ}
Let $1\leq k\leq t$ and $m=n-r$ (all the parameters are taken from Definition
\ref{defAx}). Let $\vec W_1:=\{w_1^{(1)},\ldots, w_m^{(1)}\}$\,, $\vec
W_2:=\{w_1^{(2)},\ldots, w_m^{(2)}\}$\,, $\vec Y_k:=\{y_1^{(k)},\ldots,
y_n^{(k)}\}$ and $Z_k:=\{z_1^{(k)},\ldots, z_n^{(k)}\}$ be sets of new
\emph{distinct} extension variables. The CNF formula $\neg$\emph{INJ}$_k$
consists of the following set of clauses:

\begin{enumerate}
  \item\label{itemCirEnc}

$ \norm{C^{(k)} (\vec W_1 )}(\vec Y_k);\,\,\,\,\,\norm{C^{(k)} (\vec W_2 )}(\vec Z_k)$
(expresses  that  $\vec Y_k,\vec Z_k$ are  the
 output  bits  of $C^{(k)} (\vec W_1 ),C^{(k)} (\vec W_2 )$, respectively);

\item  $ \neg u_i  \vee \neg y_i^{(k)}  \vee z_i^{(k)} ;\,\,\,\,\neg u_i  \vee y_i^{(k)}  \vee
 \neg z_i^{(k)}$, for all $1 \le i \le n$
 (expresses that $u_i $ implies $y_i^{(k)}  \equiv z_i^{(k)}$);
 \label{itINJ2}

  \item
 $v_i  \vee w_i^{(1)}
 \vee w_i^{(2)}$; $v_i  \vee \neg w_i^{(1)}  \vee \neg w_i^{(2)}$; for al $1 \le i \le m$
 (expresses that $\neg v_i $ implies $w_i^{(1)}  \equiv w_i^{^{(2)} } $);
\label{itINJ3}

\item $u_1 , \ldots ,u_n$ (expresses  that $\vec Y \equiv \vec Z$);
	\label{itINJ4}

\item $\neg v_1  \vee  \ldots  \vee \neg v_m$ (expresses  that $\vec W_1
  \not\equiv \vec W_2$);   \label{itINJ5}

\end{enumerate}
%
%
%
%
%
\end{definition}

For simplicity of writing we introduce the following notation: Let $\ell$ be a
literal and let $A$ be a CNF formula. We denote by $\ell\circvee A$ the set of
clauses (that is, the CNF formula) that results by adding to each clause of $A$
the literal $\ell$.

We now encode as a CNF formula denoted by  $\neg$INJ the negation of $${\bigwedge_{k = 1}^t
{\left( {C^{(k)} (\vec W_1 ) \equiv C^{(k)} (\vec W_2 )\, \to \vec W_1 \equiv \vec W_2 }
\right)} }.$$

\begin{definition}[{$\neg$INJ}]
The \emph{CNF} formula $\neg$\emph{INJ} consists of the following
set of clauses:
\begin{enumerate}
\item $\neg p_k  \circvee \neg {\rm{INJ}}_k$ for al $1 \le k \le
t$ (expresses that INJ$_k$ implies $\neg p_k$);

\item $p_1  \vee  \ldots  \vee p_t $ (expresses $\vee_{{k = 1}}^t {\neg {\rm{INJ}}_k }$.)

\end{enumerate}
\end{definition}

In a similar manner one can encode as a CNF the negation of the formula
$${\bigwedge_{1 \le i < j \le t} {\left( {C^{(i)} (\vec W_1 ) \not\equiv
C^{(j)} (\vec W_2 )} \right)}},$$ denoted by $\neg$DSJ (where DSJ stands for
\emph{disjoint}). We shall not develop the encoding precisely, as this is
pretty much similar to $\neg$INJ.

The last part of the promise axiom we need to encode is the formula
$$\bigvee_{i = 1}^t {C^{(i)} (\vec W_1 ) \equiv X}.$$ We denote the CNF
encoding of this formula by RST (which stands for \emph{restriction}). Again,
this is similar to the encoding of $\neg$INJ, but we show how to encode it
anyway, since we would like to illustrate in the sequel how resolution can use
RST to efficiently prove some basic facts about the $X$ variables (in the case
the circuits in $C$ have certain simple form).

\begin{definition}[{RST}]\label{defRST}
For every $1\leq k\leq t$, recall that $\vec Y_k:=\{y_1^{(k)},\ldots,
y_n^{(k)}\}$ are the output variables of $\norm{C^{(k)} (\vec W_1 )}$ from
Definition~\ref{defINJ}. The \emph{CNF} formula \emph{RST} consists of the
following set of clauses:

\begin{enumerate}
\item
$ \neg f_i^{(k)}  \vee \neg y_i^{(k)}  \vee x_i ;\,\,\,\,\neg f_i^{(k)}  \vee
y_i^{(k)}  \vee \neg x_i$ for all $1 \le i \le n$
(expresses that $f_i^{(k)}$ implies $y_i^{(k)}  \equiv x_i$);

\item $ \neg h_k  \vee f_1^{(k)} , \ldots ,\neg h_k \vee f_n^{(k)}$ (expresses that $h_k$ implies $\vec Y_k  \equiv X$);

\item  $h_1  \vee  \ldots  \vee h_t \,\,\,\,
{\rm{(expresses}}\,\,\,\bigvee\limits_{i = 1}^t {\vec Y_k  \equiv X}$.)

\end{enumerate}

\end{definition}

Finally, the promise axiom PRM$_{C,\Lambda}$ is the following CNF formula:

\begin{definition}[{CNF encoding of PRM$_{C,\Lambda}$}]
\label{defEncAx}
The CNF encoding of the promise axiom PRM$_{C,\Lambda}$ consists of the
following clauses:

\begin{enumerate}
\item $\neg q_1  \circvee \neg {\rm{INJ}}$
(expresses that {INJ} implies $\neg q_1$);\label{itINJ}

\item $\neg q_2  \circvee \neg {\rm{DSJ}}$
(expresses that {DSJ} implies $\neg q_2$);

\item $q_1  \circvee (q_2  \circvee {\rm{RST}})$ (expresses $\neg
{\rm{INJ}} \vee \neg {\rm{DSJ}} \vee {\rm{RST}}$,
which is equivalent to ${\rm{INJ}} \wedge {\rm{DSJ)}} \to
{\rm{RST}}\,{\rm{).}}$ \label{itRST}

\end{enumerate}
\end{definition}

\subsubsection{Proving basic facts about encoded circuits inside resolution}
\label{secInsideRes}

The following simple claim illustrates how one can reason inside resolution,
and specifically can ``eliminate implications'' inside resolution.
Consider, for instance, line
\ref{itINJ} in PRM$_{C,\Lambda}$ (Definition~\ref{defEncAx}).
This line expresses that INJ implies $\neg q_1$.
In other words, it is logically equivalent to $\mathrm{INJ}\to \neg q_1$.
Assume that we already know INJ (which formally means that we have a resolution
refutation of $\neg$INJ). We would like to arrive inside resolution
at $\neg q_1$.
The following straightforward claim illustrates how to do this in resolution.

\begin{claim}\label{claSimp}
\label{lemBox} Let $A$ be an unsatisfiable CNF formula with a resolution
refutation of size $s$ and let $\ell$ be any literal. Then there is a
resolution proof of $\,\ell$ from $\ell\circvee A$ of size $s$.
\end{claim}

\begin{proof}
Assume that the resolution refutation of $A$ is the sequence of clauses
$A_1,\ldots, A_s$, where $A_s=\Box$ (the empty clause). Then $\ell\vee
A_1,\ldots, \ell\vee A_s$ is a resolution proof of $\ell\vee\Box=\ell$ from
$\ell\circvee A$ (we assume that $\ell$ is not in any $A_i$; or else, by the
weakening rule, the claim also holds).
\end{proof}

Note that Claim~\ref{claSimp} implies that if there is a refutation
of $\neg$INJ of size
$s$, then there is also a proof of $\neg q_1$ of the same size $s$, from line
\ref{itINJ} in PRM$_{C,\Lambda}$ (Definition~\ref{defEncAx}).

We now illustrate how resolution can efficiently prove a certain simple fact
about simple circuits. This is needed (among other efficient proofs of similar
simple facts) in order to show the upper bound in Section~\ref{secBigPro} (and
specifically, it is used in Claim~\ref{claProvability}). Other similar facts
about the Boolean circuits constructed in Section~\ref{secBigPro} can be proved
inside resolution in a similar manner, and we shall not describe these proofs
here.


For some $1\leq k\leq t$, let $C^{(k)}$ be a circuit from a sequence of
circuits $C$ (as in the promise axioms), where $m$ and $n$ are the number of
input and output variables of $C^{(k)}$, respectively. Assume that the $i$th
output bit of $C^{(k)}$ computes the $j$th input bit $w_j$ for some $1\leq
j\leq m$ and $1\leq i\leq n$. We require that resolution can efficiently refute
(the encoding via Definition~\ref{defINJ} of):

$$C^{(k)} (\vec W_1 ) \equiv C^{(k)} (\vec W_2 )\, \wedge  w_j^{(1)}\not\equiv
w_j^{(2)}$$ (note that by assumption this is clearly a contradiction).

Since $C^{(k)}_i$ just computes the $j$th input bit $w_j$, then in fact we can
assume that the encoding $\|C^{(k)}_i(\vec W)\|(y_i)$ consists of only the
single clause $w_j$ (remember that by Definition~\ref{defEnc} we identify
between the variable \emph{encoding an input gate} with the input variable
$w_j$ itself; and here we know that $w_j$ is also the output variable). Thus,
by~\ref{itINJ2} in Definition~\ref{defINJ} we have that the output bit $y_j$
of $C^{(k)}_i(\vec W_1)$ equals the output bit $z_j$ of $C^{(k)}_i(\vec W_2)$,
and $y_j$ is actually $w^{(1)}_j$ and $z_j$ is actually $w^{(2)}_j$. Therefore,
by Definition~\ref{defINJ}~\ref{itINJ3}, we can prove $v_j$. So, by one
resolution rule applied to~\ref{defINJ}~\ref{itINJ5}, we are left with
$\vee_{i\neq j} v_i$.

Assume that all but a constant number of the output bits of $C^{(k)}$ compute
some (distinct) input bit $w_j$, for some $1\leq j\leq m$ (this assumption
corresponds to the circuits we build in Section~\ref{secBigPro}). Then the
process described in the previous paragraphs can be iterated for all such
output bits of $C^{(k)}$, in order to cut off (that is, resolve over) all the
$v_j$ variables in clause~\ref{itINJ5}  in Definition~\ref{defINJ}, until we
reach only a disjunction of \emph{constant number} of variables $v_j$ instead
of clause~\ref{itINJ5} in~\ref{defINJ}.

We are thus left with a \emph{constant number} of circuits depending only on a
\emph{constant number} of input variables. Therefore, we can now refute with a
polynomial-size resolution refutation the encoding of
\begin{equation}\label{eqIllustrate}
C^{(k)} (\vec W_1 ) \equiv C^{(k)} (\vec W_2 )\, \wedge  \vec W_1\not\equiv
\vec W_2
\end{equation}
(if indeed the circuit $C^{(k)}$ computes an injective map, which means that
 (\ref{eqIllustrate}) is unsatisfiable).

\subsubsection{Comments on decoding the encoded promise axioms}

In order to assert that promise resolution is a Cook-Reckhow proof system (see
the first paragraph in Section~\ref{secBG} for a definition) we need to make
sure that a promise resolution refutation can be identified as such in
polynomial-time. For this, one needs to be able to verify whether a given CNF
is an instance of the promise axiom.

As mentioned in Section~\ref{secMod}, this can be done by ``decoding" the CNF
that encodes the promise axiom PRM$_{C,\Lambda}$ and then checking that each
circuit in $C$ has the right number of input and output bits. Here we
illustrate how this can be achieved.

First, it is possible to identify which are the clauses pertaining to the
promise axioms out of all the clauses in the refutation (for instance, any
clause used as an axiom that is not one of the clauses of the CNF meant to be
refuted). Second, it is possible to identify which are the clauses of the
promise axiom that are part of the circuit encoding (that is, clauses in line
\ref{itemCirEnc} in Definition~\ref{defINJ}). It is then possible to decode the
circuits from the encoding, and check that the circuits are legitimate ones and
have the intended number of input and output variables (we omit the details).

\section*{Acknowledgments}
The second author is indebted to Ran Raz for very helpful conversations that
led to the present paper. We wish to thank Jan Kraj{\'{\i}}{\v{c}}ek for
commenting on an early version of this paper and Eli Ben-Sasson and Amnon
Ta-Shma for useful correspondence and conversations.

\end{document}